\documentclass[12pt]{article}
\pdfoutput=1

\usepackage[utf8]{inputenc}
\usepackage[left=2.55cm, right=2.55cm, top=2.55cm, bottom=2.55cm]{geometry}
\usepackage{amsmath,amssymb,amsbsy}
\usepackage{slashed}
\usepackage{xcolor}
\usepackage{graphicx}
\usepackage{url}
\usepackage{cancel}
\usepackage{cite}
\usepackage[colorlinks=true,allcolors=blue,pdfborder={0 0 0},linktocpage=false,pdfencoding=auto]{hyperref}
\usepackage{tabularx,booktabs}
\usepackage{multicol}
\usepackage{feynmp}
\usepackage{units}
\usepackage{xspace}
\usepackage[labelfont=bf]{caption}
\usepackage[section]{placeins}
\usepackage{subcaption}
\usepackage{soul}
\usepackage{bbold}
\usepackage{parskip}
\usepackage{float}
\usepackage{tabulary}

\definecolor{darkred}{rgb}{0.6,0,0}
\definecolor{darkpurple}{rgb}{0.5,0,0.5}

\newcommand{\beqn}{\begin{eqnarray}}
\newcommand{\eeqn}{\end{eqnarray}}
\def\non{\nonumber\\}

\begin{document}

\author{Amin Aboubrahim$^a$\footnote{\href{mailto:aabouibr@uni-muenster.de}{aabouibr@uni-muenster.de}}~, Pran Nath$^b$\footnote{\href{mailto:p.nath@northeastern.edu}{p.nath@northeastern.edu}}~ and Zhu-Yao Wang$^b$\footnote{\href{mailto:wang.zhu@northeastern.edu}{wang.zhu@northeastern.edu}} \\~\\
$^{a}$\textit{\normalsize Institut f\"ur Theoretische Physik, Westf\"alische Wilhelms-Universit\"at M\"unster,} \\
\textit{\normalsize Wilhelm-Klemm-Stra{\ss}e 9, 48149 M\"unster, Germany} \\
$^{b}$\textit{\normalsize Department of Physics, Northeastern University, Boston, MA 02115-5000, USA} \\
}

\title{\vspace{-2cm}\begin{flushright}
{\small MS-TP-21-27}
\end{flushright}
\vspace{1cm}
\Large \bf
A cosmologically consistent millicharged dark matter solution to the EDGES anomaly of possible string theory origin
 \vspace{0.5cm}}

\date{}
\maketitle

\begin{abstract}

Analysis of EDGES data shows an absorption  signal of the redshifted 21-cm line of atomic hydrogen at $z\sim 17$ which is stronger than
expected from the standard $\Lambda$CDM model. The absorption signal interpreted as brightness temperature $T_{21}$ 
 of the 21-cm line gives
an amplitude of $-$500$_{-500}^{+200}$ mK
at 99\% C.L. which is a 3.8$\sigma$ deviation from what the standard $\Lambda$CDM cosmology gives. We present a particle physics model for the baryon cooling
where a fraction of the dark matter resides in the hidden sector with a $U(1)$ gauge
symmetry and  a Stueckelberg mechanism operates mixing the visible and the hidden sectors with the hidden sector consisting of dark
Dirac fermions and dark photons.
The Stueckelberg mass mixing mechanism automatically generates a millicharge for the hidden sector dark fermions providing a theoretical basis for using millicharged  dark matter to produce the desired cooling of baryons seen by EDGES by scattering from millicharged dark matter.  We compute the relic density of the millicharged dark matter
by solving a set of coupled equations for the dark fermion and dark photon yields and for the temperature ratio of the hidden sector and the visible sector heat baths. For the analysis of baryon cooling, we analyze the evolution equations for
the temperatures of baryons and millicharged dark matter as a function of the redshift.
We exhibit regions of the parameter space which allow consistency with the
EDGES data. We note that  the Stueckelberg mechanism  arises naturally in strings and the existence of a millicharge would
point to its string origin.

\end{abstract}

\numberwithin{equation}{section}

\newpage

{  \hrule height 0.4mm \hypersetup{colorlinks=black,linktocpage=true} \tableofcontents
\vspace{0.5cm}
 \hrule height 0.4mm}

\section{Introduction}

The 21-cm line plays an important role in analysis of physics during the dark ages and the cosmic dawn in
the evolution of the early 
universe~\cite{Pritchard:2011xb,Morales:2009gs,Furlanetto:2006jb,Munoz:2018pzp,Fialkov:2018xre,Barkana:2018qrx,Pritchard:2008da,Loeb:2003ya}.
That line arises from the spin transition from the triplet  state to the singlet state
and vice-versa in the ground state of neutral hydrogen. The relative abundance of the triplet and
the singlet states defines the spin temperature of the hydrogen gas and is given by~\cite{Munoz:2015bca}
\begin{align}
\frac{n_1}{n_0} = 3 e^{-\frac{T_*}{T_s}},
\end{align}
where 3 is the ratio of the spin degrees of freedom for the triplet versus the singlet state, $T_*$ is defined
by $\Delta E=k T_*$, where $\Delta E= 1420$ MHz 
is the energy difference at rest between the two spin states, $T_*= 0.068$K, and $T_s$ is the spin temperature of the hydrogen gas.
The difference $T_s-T_\gamma$ between the spin temperature and the background
temperature is an important avenue for the
exploration of the dark period in the early history of the universe.  In an expanding universe, the 21-cm line is redshifted according to the formula
  $\nu= 1420/(1+z)$ MHz, where $z$ is the redshift. Cohen \textit{et al}~\cite{Cohen:2017xpx} analyzed the 21-cm global signal as a function of the redshift and found that in the standard cosmology the temperature of that line, $T_{21}$, can be 
  maximally $T_{21} \simeq - 250$ mK for redshifts in the range $z=($6$-$40). 
The Experiment to Detect the Global Epoch of Reionization Signature (EDGES) reported an absorption profile centered at the frequency $\nu=78$ MHz in the sky-averaged spectrum~\cite{Bowman:2018yin}. Bowman \textit{et al}~\cite{Bowman:2018yin} parameterize the intensity of the observed 21-cm signal by $T_{21}(z)$ defined as 
\begin{align}
T_{21}(z) \simeq 0.023\,x_{\rm HI}(z)
 \left[\frac{0.15}{\Omega_m}\frac{(1+z)}{10}\right]^{\frac{1}{2}}\left(\frac{\Omega_B h^2}{0.02}\right) 
\left[1-\frac{T_\gamma(z)}{ T_s(z)} \right] \text{K}. 
\label{t21z}
\end{align}
Here $x_{\rm HI}$ is the fraction of neutral hydrogen, $\Omega_{\rm m}$ and $\Omega_{\rm B}$ are fractions of the critical density for  matter and for  baryons, $h$ is the Hubble parameter in units of  100 km/s/Mpc,
and $T_\gamma(z)$ is the photon temperature at redshift $z$. The analysis of Ref.~\cite{Bowman:2018yin} finds $T_{21} = - 500_{-500}^{+200}$ mK
at 99\% C.L. as compared to the $\Lambda$CDM model of $-230$ mK,  which shows that the
EDGES result is a 3.8$\sigma$ deviation from that of
the standard cosmological paradigm. As noted above, the signal is important in that it can provide information
about the early phase of cosmic structure formation. 
Subsequently, Barkana~\cite{Barkana:2018lgd} (based on~\cite{Munoz:2015bca}) pointed out quite generally that the observed effect could arise from a smaller than expected  temperature difference between baryon and photons. The work of~\cite{Barkana:2018lgd} further showed that cooling down the baryons by roughly 3 K would explain the observation. In fact, several mechanisms have been
proposed to explain the 3.8$\sigma$ anomaly~\cite{Munoz:2018pzp,Munoz:2018jwq,Halder:2021jiv,Liu:2019knx,Berlin:2018sjs,Feng:2018rje,Barkana:2018lgd,Barkana:2018cct,Fraser:2018acy,Pospelov:2018kdh,Moroi:2018vci,Fialkov:2019vnb,Choi:2019jwx,Creque-Sarbinowski:2019mcm,Kovetz:2018zan,Bondarenko:2020moh,Lanfranchi:2020crw}.
These include the possibility, as noted above,  that the baryons are cooler than what $\Lambda$CDM predicts but also the possibility that the
CMB  background radiation temperature was hotter than expected. Additionally, possible causes could be astrophysical phenomena such as radiation from stars
and star remnants~\cite{Bowman:2018yin,Barkana:2018lgd} though these are deemed less likely.
Regarding the first possibility, i.e., the cooling of baryons,  one specific proposal is the
existence of a small percentage of dark matter which is millicharged  and can generate the desired absorption amplitude seen by EDGES~\cite{Munoz:2018pzp,Munoz:2018jwq,Halder:2021jiv,Liu:2019knx,Berlin:2018sjs}. Some issues regarding the millicharge solution were raised in~\cite{Barkana:2018cct,Berlin:2018sjs} which, however, can be overcome~\cite{Liu:2019knx}. 

We note here that some concerns have been raised~\cite{Hills:2018vyr} regarding the conclusion of~\cite{Bowman:2018yin}
 that the appearance of the 21-cm absorption line on the microwave background is arising due to the effect of light from the first stars on the hydrogen 
atom.  Thus, the  analysis of~\cite{Hills:2018vyr}  indicates that the evidence does
not necessarily rise to the level for invoking new physics, although their work does not prove that the 21-cm signal is absent.
In the analysis below we assume  the existence of the EDGES signal. However,  we believe more  data is certainly needed to confirm the signal
at the level needed to claim discovery of new physics.

In this work we will focus on the further exploration of cooling of the baryons by
millicharged dark matter. Of course the use of millicharged dark matter begs the question: how did the
millicharged dark matter originate in the first place?
We address this question in this work.
Specifically we wish to investigate the EDGES effect  within a well defined particle physics model
 which automatically generates a millicharge through mass mixing, where we compute the relic density of such millicharged dark matter
along with the evolution of the dark matter and baryon temperatures. Specifically we consider a hidden sector with an extra $U(1)$ gauge symmetry beyond the gauge group of
the Standard Model, where the hidden sector contains a dark fermion $D$ which has normal strength interaction with a dark photon $\gamma'$ and feeble interactions with the visible sector.
The analysis is done within the framework of the Stueckelberg mechanism which
generates a millicharge for the dark fermion which constitutes the millicharged DM in our model.
We analyze the millicharged DM-baryon interactions  which affect the thermal history after recombination and cools the baryon fluid enough to produce the EDGES signal. The millicharged DM model is subject to
  stringent constraints from big-bang nucleosynthesis (BBN)~\cite{Boehm:2013jpa}, stellar cooling, Supernova 1987A (SN1987A)~\cite{Chang:2018rso}, and SLAC~\cite{Prinz:1998ua} millicharge
experiment as well as CMB limits from the Planck 2015 data~\cite{Boddy:2018wzy}. We show that within these constraints as well as the constraint on $\Delta N_{\rm eff}$
the proposed model produces the desired absorption amplitude of the 21-cm line observed by
 EDGES. 

 The  organization of  the rest of the paper as follows:
In section~\ref{sec:model} we give the particle physics model used to discuss the  EDGES effect.
In section~\ref{sec:early} we discuss the evolution equations for the yields of the dark fermion and the dark photon
 as a function of the hidden sector temperature. Here we also give the evolution equation for the ratio of
 temperatures of
 the visible sector and the hidden sector, and discuss the computation of the relic density  of
 millicharged dark fermions which constitute a small fraction of dark matter.
In section~\ref{sec:late}  an analysis of the coupling of millicharged dark matter and baryons responsible for baryon cooling
is given.
In section~\ref{sec:T21} we discuss the brightness temperature of the 21-cm line measured by EDGES.
A numerical analysis of the millicharged DM relic density and fit to EDGES data consistent with
all constraints is given in section~\ref{sec:numerical}, and conclusions are given in section~\ref{sec:conclu}.
Further details related to the analysis are given in appendices~\ref{app:A},~\ref{app:B} and~\ref{app:C}.

\section{A particle physics model for EDGES}\label{sec:model}

We extend the Standard Model (SM) gauge group by an extra $U(1)_X$ under which the SM is neutral. The extra gauge field $C^{\mu}$ mixes with the SM $U(1)_Y$ hypercharge $B^{\mu}$
via kinetic mixing~\cite{Holdom:1985ag}.
We further introduce a Stueckelberg mass mixing~\cite{Kors:2004dx}
 between those groups and the total Lagrangian is given by
 \begin{equation}
\mathcal{L}=\mathcal{L}_{\rm SM}+\mathcal{L}_{\rm hid}+ \mathcal{L}_{\rm SM-hid}, 
\label{totL}
\end{equation}
where $\mathcal{L}_{\rm SM}$ is the Standard Model Lagrangian,  $\mathcal{L}_{\rm hid}$ is the Lagrangian for the hidden sector and 
$\mathcal{L}_{\rm SM-hid}$ is the overlap Lagrangian connecting the visible sector to the hidden sector.  The latter and the former are given by 
\begin{align}
\label{totL1}
\mathcal{L}_{\rm hid} &=-\frac{1}{4}C_{\mu\nu}C^{\mu\nu}+i\bar D \gamma^\mu \partial_\mu D -m_D \bar D D - \frac{1}{2} 
(\partial_\mu \phi \partial^\mu \phi) + \mathcal{L}^{\rm int}_{\rm hid}, \\
 \label{totL3}
 \mathcal{L}_{\rm SM-hid} &= 
-\frac{\delta}{2}C_{\mu\nu}B^{\mu\nu}-\frac{1}{2}(\partial_{\mu}\sigma+M_1 C_{\mu}+M_2 B_{\mu})^2,
\end{align}
where
\begin{align}
\label{totL2} 
 \mathcal{L}^{\rm int}_{\rm hid}&= g_X Q_X \bar D \gamma^\mu D C_\mu + \lambda D\bar{D}\phi.
\end{align}
In Eq.~(\ref{totL1}), $D$ is the dark fermion in the hidden sector, $\phi$ is a scalar field whose role will be discussed later and in Eq.~(\ref{totL3}), $C_{\mu}$ is the Stueckelberg gauge field of  $U(1)_X$  of the hidden sector and  $\mathcal{L}^{\rm int}_{\rm hid}$ is the interaction in the hidden sector between the gauge field $C_\mu$ and the dark fermion where in Eq.~(\ref{totL2}) 
$g_X$ is the gauge coupling constant of the field $C_\mu$ and $Q_X$ is the $U(1)_X$ charge of $D$. The last term in Eq.~(\ref{totL2}) is the Yukawa coupling $\lambda$ of the scalar field with the dark fermion which 
plays a role in the evolution of the temperature ratio of the hidden to the visible sector in the very low temperature regime. 
Eq.~(\ref{totL3}) describes the overlap between the visible sector and the hidden sector where the first term on the right-hand-side gives the kinetic mixing of
 the $U(1)_Y$ hypercharge field $B_\mu$ of the Standard Model and the gauge field $C_\mu$ of the hidden sector, while the second term on the right-hand-side gives the Stueckelberg mass mixing between the two $U(1)$ fields, where $M_1$ and $M_2$ are two mass parameters. The setup of Eq.~(\ref{totL})$-$Eq.~(\ref{totL3})
 guarantees  the existence of a massless mode (the photon) when canonical diagonalization of the full Lagrangian of Eq.~(\ref{totL}) is carried out.

    In this analysis we are interested in millicharge  couplings $ \mathcal{L}^{\rm m}$
        arising from the inclusion of the kinetic and Stueckelberg mass mixing 
    between the visible sector and the hidden sector. These couplings have two sources so that 
   \begin{align}
    \mathcal{L}^{\rm m}&= \mathcal{L}_{\rm SM}^{\rm m} +   \mathcal{L}_{\rm hid} ^{\rm m},
    \label{milli-tot} 
      \end{align}
where the first term on the right-hand-side contains the millicharge couplings arising from the Standard Model and the second term gives
the millicharge couplings arising  from the  hidden sector.
Both of these enter in the analysis of this work.  
After canonical diagonalization of Eq.~(\ref{totL}), one finds that the three fields $C^\mu, B^\mu$ and $A^\mu_3$, where $A_3^\mu$ is the third component of 
the SM $SU(2)_L$ left field $A^\mu_a$ ($a=$1$-$3),  mix and  become a linear combination of the mass eigenstates  $Z_\mu, A_\mu^{\gamma'},  A_\mu^{\gamma}$,
where $Z_\mu$ is the $Z$-boson field with mass $m_Z$, $A_\mu^{\gamma'}$ is the dark photon field with mass $m_{\gamma'}$ and $A_\mu^{\gamma}$
is the photon field. The effect of canonical diagonalization of the kinetic and mass square terms of the gauge fields is to generate millicharge
couplings of two types:
\begin{enumerate}
\item
The dark fermions of the hidden sector develop millicharge coupling to the photon field and  to the $Z$-boson (see Eq.~(\ref{milli-hid1})),
while they have normal strength 
coupling to the dark photon field proportional to $g_X Q_X$ (see Eq.~(\ref{D-darkphoton})).
\item
The quarks and leptons of the Standard Model develop millicharge size couplings to the dark photon (see Eq.~(\ref{milli-sm})).
\end{enumerate}
The couplings $\mathcal{L}_{\rm SM}^{\rm m}$ will be discussed in Appendix~\ref{app:A} and here 
we  discuss $\mathcal{L}_{\rm hid}^{\rm m}$ which arise from the couplings of the dark fermion with the field $C_\mu$ so that
\begin{align}
\mathcal{L}_{D}^{\rm{\rm gauge}} &= g_X Q_X \bar D \gamma^\mu D C_\mu.
\label{nc_2} 
\end{align}
In the canonically diagonalized frame, $\mathcal{L}_{\rm hid} ^{\rm m}$ is given by 
\begin{align}
\label{milli-hid1}
\mathcal{L}_{\rm hid} ^{\rm m}
&=\bar{D}\gamma^{\mu}
                        \left[\epsilon_{Z}^{D} Z_{\mu}
                        +\epsilon_{\gamma}^{D}
                        A^{\gamma}_{\mu}\right]D.
\end{align}
The full expressions for $\epsilon_{Z}^{D}$ and  $\epsilon_{\gamma}^{D}$ are given in Appendix~\ref{app:A}. Here we 
show results in the  limit when the kinetic mixing
parameter $\delta$ and the Stueckelberg mass mixing parameter $\epsilon= M_2/M_1$ are small such that $\epsilon<<1, \delta <<1$,
which give 
\begin{align}
\label{milli-hid2}
\epsilon_{Z}^D&\simeq  e \epsilon Q_X g_X\sin\theta_W,\non
\epsilon_{\gamma}^{D}&\simeq  -e\epsilon Q_X
\frac{g_X}{g_Y},
\end{align}
where $\sin\theta_W$ is the weak angle of the Standard Model
and the millicharge is defined as $\epsilon_D= -\epsilon_{\gamma}^{D}/e$ so that 
\begin{equation}
\label{milli-hid3}
\epsilon_D=\frac{g_X Q_X}{g_Y}\epsilon\,.
\end{equation}
Thus to lowest order in $\epsilon, \delta$, the photon and the $Z$-boson have millicharge couplings to the dark
fermions which are independent of $\delta$ and depend only on $\epsilon$ which arises from the Stueckelberg mass mixing, 
 while the dark photon has normal strength coupling to the dark fermions which is given by
\begin{align}
\label{D-darkphoton}
 \mathcal{L}^{D}_{\gamma'} &= 
  g_{\gamma'}^D 
 \bar D\gamma^\mu 
  D A_{\mu}^{\gamma'},
\end{align}
where $g_{\gamma'}^D\simeq  g_X Q_X$.
  Further details are given in Appendix~\ref{app:A} and in~\cite{Aboubrahim:2020lnr,Aboubrahim:2020afx,Aboubrahim:2021ycj}
  where $\mathcal{L}_{\rm SM}^{\rm m}$ is also discussed.

We discuss now the reason the Stueckelberg extension of the Standard Model  is critical in our analysis of the EDGES signal.    
      As seen above, the Stueckelberg extension contains 
 couplings of the $D$ fermion with $A_\mu^\gamma$ and $Z$ (Eq.~(\ref{milli-hid1})),
    couplings of the $D$ fermion with $A_\mu^{\gamma'}$ (Eq.~(\ref{D-darkphoton})), couplings of Standard Model quark and leptons with $A_\mu^{\gamma'}$
    (Eq.~(\ref{milli-sm})). All of these couplings enter in our analysis of the EDGES signal.
     In the absence of the Stueckelberg couplings only the $A_\mu^\gamma$ term in Eq.~(\ref{milli-hid1}) will exist,  and the $Z$ boson coupling
    in  Eq.~(\ref{milli-hid1}) will be absent, and all of the couplings of Eq.~(\ref{D-darkphoton}) and of Eq.~(\ref{milli-sm}) will disappear. 
      In this case it would not be possible to carry out a two-temperature evolution of the visible and hidden sectors nor 
   obtain a consistent cosmology discussed in the following sections. 
  Further, we note that if millicharged particles exist, a natural
 setting for their existence points to strings and we are not aware of any alternative mechanism that does that. 
  This is so because the Stueckelberg mechanism is intimately tied~\cite{Kors:2004iz}   
 to the Green-Schwarz anomaly cancellation in 
 string theory~\cite{Green:1984sg,Green:1987sp}. The existence of millicharged particles was predicted in the Stueckelberg extension of the standard model with mass mixing in \cite{Kors:2004dx}.

\section{Temperature evolution at early times}\label{sec:early}

  Typically the cosmological analyses often bifurcate into the early times or the large redshift era defined by $z\gg 3000$
  and late times of the small redshift era defined by its compliment $z\ll3000$.
Here we will discuss the relevant dynamical equations for early times and in the subsequent section we will discuss analyses at
late times and specially for $z$ in the vicinity of where the effect observed by EDGES occurs.
As noted in the section above we are dealing with two sectors one visible and the other hidden. As pointed out in several recent works~\cite{Aboubrahim:2020lnr,Aboubrahim:2021ycj,Aboubrahim:2021dei,Foot:2014uba,Foot:2016wvj}, in such a situation, an accurate analysis requires that one deals with the visible and the hidden sectors baths being at different temperatures,
 with the visible sector at temperature $T$ and the hidden sector at temperature $T_h$. The temperature difference between
 the two sectors has a very significant effect on the analysis as is easy to see since the quantities which enter in the cosmological
 analyses such as the entropy density $s$, energy density $\rho$ and pressure density $p$ have direct dependence on temperature,
 i.e, $s\sim T^3, (\rho, p) \sim T^4$.  In the analysis here we will take account of the visible and the hidden sectors in different heat baths.
 Thus the  Boltzmann equations for  the number densities $n_D$ for the dark fermion, $n_{\gamma'}$  for the dark photon {and $n_\phi$ for the scalar depend on the  two temperatures $T$ and $T_h$ because of coupling between the hidden sector 
and the visible sector.

The above implies that a determination of the
relic density of dark fermions, which we assume to have feeble interactions with the visible sector, will
involve three coupled equations for $dn_D/dt$,  $dn_{\gamma'}/dt$, $dn_\phi/dt$  and $d\eta/dt$, where $\eta=T/T_h$.  In the analysis we will use the constraint that the entropies of the hidden sector and of the visible sector are not individually conserved but it is only the total entropy that is conserved,
i.e., $S=\mathbb{s} R^3$ is conserved where $S$ is the sum of the visible and the hidden sector entropies. This leads to the evolution equation for
the entropy density $\mathbb{s}$ so that  $d\mathbb{s}/dt + 3 H\mathbb{s}=0$, where $H$ is the Hubble parameter and
  $\mathbb{s}=\mathbb{s}_v+ \mathbb{s}_h$, where $\mathbb{s}_v$ depends on $T$ and $\mathbb{s}_h$ on $T_h$ so that
\begin{align}
\mathbb{s}&=\frac{2\pi^2}{45}\left(h_{\rm eff}^h T_h^3+h_{\rm eff}^v T^3\right).
\label{eq3.a}
\end{align}
Here $h^v_{\rm eff} (h^h_{\rm eff})$ is the visible(hidden) effective entropy degrees of freedom.

 The Hubble parameter also depends on both $T$ and $T_h$ as can be seen
from the Friedman equation
\begin{align}
H^2= \frac{8\pi G_N}{3} (\rho_v(T)  +\rho_h(T_h)),
\label{eq3.b}
\end{align}
where $\rho_v(T) (\rho_h(T_h))$ is the energy density in the visible (hidden) sector at temperature $T(T_h)$ and given by
\begin{align}
\rho_v&=\frac{\pi^2}{30}g_{\rm eff}^v T^4, ~~
\rho_h=\frac{\pi^2}{30}g_{\rm eff}^h T_h^4.
\label{rho-1}
\end{align}
The visible effective entropy degrees of freedom $h^v_{\rm eff}$ in Eq.~(\ref{eq3.a}), and the visible  effective energy degrees of freedom 
$g^v_{\rm eff}$ in Eq.~(\ref{rho-1}) are standard and we use tabulated results from lattice QCD~\cite{Drees:2015exa}.
Here we focus on  $h^h_{\rm eff}$ in Eq.~(\ref{eq3.a}) and on $g^h_{\rm eff}$  in Eq.~(\ref{rho-1}) which are
the entropy and energy density degrees of freedom for 
the hidden sector. They include degrees of freedom for the  dark photon, the dark fermion and the massless scalar $(g_\phi=h_\phi=1)$ so that
\begin{align}
g^h_{\rm eff}&= g^{\gamma'}_{\rm eff} +\frac{7}{8}  g^D_{\rm eff}+g_\phi,~~~\text{and}~~~h^h_{\rm eff}= h^{\gamma'}_{\rm eff} + \frac{7}{8} h^D_{\rm eff}+h_\phi,
\end{align}
where at  temperature $T_h$, $g_{\rm eff}$ and  $h_{\rm eff}$  for the particles $\gamma'$ and $D$ are 
given by~\cite{Hindmarsh:2005ix} 
\begin{equation}
\begin{aligned}
g^{\gamma'}_{\rm eff}& = \frac{45}{\pi^4} \int_{x_{\gamma'}}^{\infty} \frac{\sqrt{x^2-x_{\gamma'}^2} }{e^x-1 } x^2 dx,~~~\text{and}~~~h^{\gamma'}_{\rm eff}= \frac{45}{4\pi^4} \int_{x_{\gamma'}}^{\infty} \frac{\sqrt{x^2-x_{\gamma'}^2} }{e^x-1 } 
(4x^2-x_{\gamma'}^2) dx, \\
g^{D}_{\rm eff}& = \frac{60}{\pi^4} \int_{x_{D}}^{\infty} \frac{\sqrt{x^2-x_{D}^2} }{e^x+1 } x^2 dx,~~~\text{and}~~~h^{D}_{\rm eff}= \frac{15}{\pi^4} \int_{x_{D}}^{\infty} \frac{\sqrt{x^2-x_{D}^2} }{e^x+1 } 
(4x^2-x_D^2) dx.
\end{aligned}
\label{hdof}
\end{equation}
Here $x_{\gamma'}$ and $x_D$ are defined so that $x_{\gamma'}= m_{\gamma'}/T_h$ and $x_D= m_D/T_h$. 
The  limit 
$x_{\gamma'}\to 0$  gives $g^{\gamma'}_{\rm eff}= h^{\gamma'}_{\rm eff}\to 3$ and the limit $x_D\to 0$ gives $g^{D}_{\rm eff}= h^{D}_{\rm eff}\to 4$.

 The  time evolution of $\rho_h$ is  given by             
\begin{align}
\frac{d\rho_h}{dt} +3H (\rho_h +p_h) =j_h,
\end{align}
where $p_h$ is the pressure density for the hidden sector  and $j_h$ is the source in the hidden sector and is given below in
Eq.~(\ref{y4}).

 We will use $T_h$ as the reference temperature and replace $t$ by $T_h$ and analyze the evolution of
 $n_D$, $n_{\gamma'}$, $n_\phi$ and $\eta$ as a function of $T_h$. For the computation of the relic densities,
 it is more convenient to deal directly with yields
  defined by $Y_a = n_a/\mathbb{s}$ for a particle species $a$ with number density $n_a$.
  We assume that the dark particles $D, \gamma'$ and $\phi$  are feeble and there is no initial abundance and that $D, \gamma'$ are initially
  produced only via freeze-in processes such as $i~\bar{i}\to D\bar D$, $i~\bar{i}\to \gamma'$, where $i$ refers to Standard Model
  particles. However, $D$ and $\gamma'$ have interactions such as $D\bar D\to \gamma' \gamma'$
  within the hidden sector which, in our case, are not feeble. Furthermore, $\phi$ will be produced by hidden sector processes such as $D\bar{D}\to\phi\phi$ and $D\bar{D}\to\phi\gamma'$ since it does not have a coupling with the SM.
The Boltzmann equations
  for the yields $Y_D$, $Y_{\gamma'}$ and $Y_\phi$ and the differential equation for $\eta$ then take the form
\begin{align}
\label{y1}
\frac{dY_D}{dT_h}=&-\frac{\mathbb{s}}{H}\Big(\frac{d\rho_h/dT_h}{4\zeta\rho_h-j_h/H}\Big)\Bigg[\langle\sigma v\rangle_{D\bar{D}\to i\bar{i}}(T) Y_D^{\rm eq}(T)^2 \nonumber \\
&\hspace{4cm}-\frac{1}{2}\langle\sigma v\rangle_{D\bar{D}\to\gamma'\gamma'}(T_h)\left(Y^2_D-Y_D^{\rm eq}(T_h)^2\frac{Y^2_{\gamma'}}{Y^{\rm eq}_{\gamma'}(T_h)^2}\right) \nonumber \\
&\hspace{4cm}-\frac{1}{2}\langle\sigma v\rangle_{D\bar{D}\to\phi\gamma}(T_h)\left(Y^2_D-Y_D^{\rm eq}(T_h)^2\frac{Y_{\phi}}{Y^{\rm eq}_{\phi}(T_h)}\right) \nonumber \\
&\hspace{4cm}-\frac{1}{2}\langle\sigma v\rangle_{D\bar{D}\to\phi\gamma'}(T_h)\left(Y^2_D-Y_D^{\rm eq}(T_h)^2\frac{Y_{\phi}Y_{\gamma'}}{Y^{\rm eq}_{\phi}(T_h)Y^{\rm eq}_{\gamma'}(T_h)}\right) \nonumber \\
&\hspace{4cm}-\frac{1}{2}\langle\sigma v\rangle_{D\bar{D}\to\gamma'}(T_h)\,Y^2_D+\frac{1}{\mathbb{s}}\langle\Gamma_{\gamma'\to D\bar{D}}\rangle(T_h)\,Y_{\gamma'}\Bigg],
\end{align}
\begin{align}
\label{y2}
\frac{dY_{\gamma'}}{dT_h}=&-\frac{\mathbb{s}}{H}\left(\frac{d\rho_h/dT_h}{4\zeta\rho_h-j_h/H}\right)\Bigg[-\langle\sigma v\rangle_{\gamma'\gamma'\to D\bar{D}}(T_h)\left(Y^2_{\gamma'}-Y^{\rm eq}_{\gamma'}(T_h)^2\frac{Y^2_D}{Y^{\rm eq}_D(T_h)^2}\right) \nonumber \\
&\hspace{4cm}+\frac{1}{2}\langle\sigma v\rangle_{D\bar{D}\to\phi\gamma'}(T_h)\left(Y^2_D-Y_D^{\rm eq}(T_h)^2\frac{Y_{\phi}Y_{\gamma'}}{Y^{\rm eq}_{\phi}(T_h)Y^{\rm eq}_{\gamma'}(T_h)}\right) \nonumber \\
&\hspace{4cm}-\langle\sigma v\rangle_{\gamma' D\to\gamma D}(T_h)\left(Y_{\gamma'}-Y_{\gamma'}^{\rm eq}(T_h)\right)Y_D  \nonumber \\
&\hspace{4cm}-\frac{1}{\mathbb{s}}\langle\Gamma_{\gamma'\to i\bar{i}}\rangle(T_h)\,Y_{\gamma'}+\langle\sigma v\rangle_{i\bar{i}\to\gamma'}(T)Y_i^{\rm eq}(T)^2 \nonumber \\
&\hspace{4cm}-\frac{1}{\mathbb{s}}\langle\Gamma_{\gamma'\to D\bar{D}}\rangle(T_h)\left(Y_{\gamma'}-Y^{\rm eq}_{\gamma'}(T_h)\frac{Y^2_D}{Y^{\rm eq}_D(T_h)^2}\right)\Bigg],
\end{align}
\begin{align}
\label{yphi}
\frac{dY_{\phi}}{dT_h}=&-\frac{\mathbb{s}}{H}\left(\frac{d\rho_h/dT_h}{4\zeta\rho_h-j_h/H}\right)\Bigg[\frac{1}{2}\langle\sigma v\rangle_{D\bar{D}\to\phi\gamma}(T_h)\left(Y^2_{D}-Y^{\rm eq}_{D}(T_h)^2\frac{Y_\phi}{Y^{\rm eq}_\phi(T_h)}\right) \nonumber \\
&\hspace{4cm}+\frac{1}{2}\langle\sigma v\rangle_{D\bar{D}\to\phi\gamma'}(T_h)\left(Y^2_D-Y_D^{\rm eq}(T_h)^2\frac{Y_{\phi}Y_{\gamma'}}{Y^{\rm eq}_{\phi}(T_h)Y^{\rm eq}_{\gamma'}(T_h)}\right) \nonumber \\
&\hspace{4cm}-\langle\sigma v\rangle_{\phi D\to\gamma D}(T_h)(Y_\phi-Y^{\rm eq}_\phi(T_h))Y_D \nonumber \\
&\hspace{4cm}+\frac{1}{2}\langle\sigma v\rangle_{D\bar{D}\to\phi\phi}(T_h)\left(Y^2_D-Y_D^{\rm eq}(T_h)^2\frac{Y_{\phi}^2}{Y^{\rm eq}_{\phi}(T_h)^2}\right)\Bigg], \\
& \nonumber \\
\frac{d\eta}{dT_h}= &-\frac{\eta}{T_h} +  \left[\frac{\zeta \rho_v+ \rho_h( \zeta-\zeta_h)+ j_h/(4H)}{\zeta_h\rho_h- j_h/(4H)}\right] \frac{d\rho_h/dT_h}{T_h (d\rho_v/dT)},
\label{y3}
\end{align} 
where the source term is
\begin{align}
\label{y4}
j_h=&\sum_i \Big[2Y^{\rm eq}_i(T)^2 J(i~\bar{i}\to D\bar{D})(T)+Y^{\rm eq}_i(T)^2 J(i~\bar{i}\to \gamma')(T)\Big]\mathbb{s}^2-Y_{\gamma'}J(\gamma'\to f\bar{f})(T_h)\mathbb{s} \nonumber \\
&-\left[\frac{1}{2}Y_D^2\,J(D\bar{D}\to\phi\gamma)(T_h)+Y_D Y_{\gamma'}J(\gamma' D\to\gamma D)(T_h)+Y_\phi Y_D\,J(\phi D\to\gamma D)(T_h)\right]\mathbb{s}^2,
\end{align} 
and the equilibrium yield is given by
\begin{equation}
Y^{\rm eq}_i=\frac{n_i^{\rm eq}}{\mathbb{s}}=\frac{g_i}{2\pi^2 \mathbb{s}}m_i^2 T K_2(m_i/T).
\end{equation}
Here $g_i$ is the number of degrees of freedom of particle $i$ and mass $m_i$, $K_2$ is the modified Bessel function of the second kind and degree two and the $J$ functions in Eq.~(\ref{y4}) are given by
\begin{align}
n^{\rm eq}_i(T)^2 J(i~\bar{i}\to D\bar{D})(T)&=\frac{T}{32\pi^4}\int_{s_0}^{\infty}ds~\sigma_{D\bar{D}\to i\bar{i}}s(s-s_0)K_2(\sqrt{s}/T), \\
n^{\rm eq}_i(T)^2 J(i~\bar{i}\to \gamma')(T)&=\frac{T}{32\pi^4}\int_{s_0}^{\infty}ds~\sigma_{i\bar{i}\to \gamma'}s(s-s_0)K_2(\sqrt{s}/T), \\
n_{\gamma'}J(\gamma'\to f\bar{f})(T_h)&=n_{\gamma'}m_{\gamma'}\Gamma_{\gamma'\to f\bar{f}}, \\
{J(ab\to cd)(T_h)}&{=\frac{1}{8T_h m_a^2 m_b^2 K_2(m_a/T_h) K_2(m_b/T_h)} }\nonumber \\
& {\times\int_{(m_a+m_b)^2}^{\infty} ds~\sigma_{ab\to cd}(s) s[s-(m_a+m_b)^2]K_2(\sqrt{s}/T_h).}
\end{align}
In Eq.~(\ref{y3}), $\zeta$ and $\zeta_h$ are defined so that
 $\zeta= \frac{3}{4} (1+p/\rho)$ and $\zeta_h=\frac{3}{4} (1+p_h/\rho_h)$
and interpolate between 1 and 3/4 as one transitions from radiation dominance to matter dominance. In Eq.~(\ref{y2}) there are contributions one can add on the right-hand-side which involve processes $i~\bar{i}\to \gamma' \gamma,
\gamma'Z, \gamma'\gamma'$. However, their contributions are relatively small compared to $i~\bar{i}\to \gamma'$. The thermally averaged cross sections appearing in Eqs.~(\ref{y1}) and~(\ref{y2}) are given by
\begin{equation}
\langle\sigma v\rangle_{a\bar{a}\to bc}(T)=\frac{1}{8 m^4_a T K^2_2(m_a/T)}\int_{4m_a^2}^{\infty} ds ~\sigma(s) \sqrt{s}\, (s-4m_a^2)K_1(\sqrt{s}/T),
\end{equation}
and
\begin{align}
n_i^{\rm eq}(T)^2\langle\sigma v\rangle_{i\bar{i}\to\gamma'}(T)&= \frac{T}{32\pi^4}\int_{s_0}^{\infty} ds ~\sigma(s) \sqrt{s}\, (s-s_0)K_1(\sqrt{s}/T),
\end{align}
where $K_1$ is the modified Bessel functions of the second kind and degree one and $s_0$ is the minimum of the Mandelstam variable $s$.

The relic density of $D$ is related to $Y_D$ by
\begin{align}
\Omega h^2 = \frac{m_D Y^{\infty}_D \mathbb{s}_0 h^2}{\rho_c},
\label{relic}
\end{align}
where $\rho_c$ is the critical density, $\mathbb{s}_0$ is today's entropy density, $Y^{\infty}_D$ is today's DM comoving number density and $h=0.674$\cite{Planck:2018vyg}.

The millicharged dark matter relic density  is $< 0.4\%$ of the total relic density of dark matter as given by the
Planck Collaboration~\cite{Aghanim:2018eyx} so that
$(\Omega h^2)_{\rm PLANCK}=0.1198\pm 0.0012$. This is due to the CMB constraints as discussed in several works~\cite{Munoz:2015bca,Kovetz:2018zan,Dolgov:2013una,Dubovsky:2003yn}.

\section{Millicharged dark matter and baryon coupling: temperature evolution at late times}\label{sec:late}

We now focus on the temperature evolution of the baryons and of dark matter at late times. It has been known for some time
 that the temperature evolution can be affected by Rutherford scattering of baryons and dark matter 
  and if the dark matter is at a lower temperature than the baryons, the scattering will lead to cooling of baryons.
  This situation arises if the dark matter is millicharged in which case the baryons and dark matter can scatter via
  Coulomb interactions. 
Further, 
the difference in their evolution history will produce a relative velocity between them~\cite{Munoz:2015bca} which has a significant effect in the cooling of baryons. 
In this work we focus precisely on this possibility, i.e.,  that a fraction of dark matter is 
 millicharged which couples to baryons with a cross section parameterized as  $\sigma=\sigma_0v^{-4}$ where $v$ stands for  the relative velocity between DM and baryons and $\sigma_0$ has the form~\cite{Munoz:2018pzp}
\begin{equation}
\sigma_0=\frac{2\pi\alpha^2\epsilon_D^2}{\mu_{D,t}^2}\log\left(\frac{9T_B^3}{4\pi\epsilon_D^2\alpha^3x_e n_{\rm H}}\right).
\label{sigma0}
\end{equation}
In Eq.~(\ref{sigma0}),  $T_B$ stands for the temperature of the baryons and $T_D$ stands for the temperature of DM which, at late times, becomes different from the temperature of the thermal bath maintained by $\phi$ and so $T_D\neq T_h$ (we will discuss it further in section~\ref{sec:numerical}), where $T_h$ appears in
  the Boltzmann equations, Eqs.~(\ref{y1})$-$(\ref{y3}).  Further, in  Eq.~(\ref{sigma0}),
   $\alpha=1/137$, $\epsilon_D$ is the millicharge of the DM particle defined by Eq.~(\ref{milli-hid3}), and the  quantity $\mu_{D,t}$ is the DM-target reduced
 mass (the target $t$ of mass $m_t$ where $t$ is either an electron or a proton) so that
\begin{equation}
\mu_{D,t}=\frac{m_{D}m_t}{m_{D}+m_t}.
\end{equation}
 Finally, $x_e$ is the ionization rate which is determined via the solution of the coupled equations, Eqs.~(\ref{QB})$-$(\ref{vdb}),
 and $n_H$ is the number density of hydrogen nuclei.\footnote{More general expressions for $\sigma_0$ exist in the literature, see Ref.~\cite{Liu:2019knx}. After correcting typos, we can write this equation as $\sigma^{\rm bm}=\sigma_0 v_{\rm rel}^{-4}$ so that $$\sigma_0 = \frac{2\pi Q^2\alpha^2_{\rm EM}}{\mu^2_{\rm m}}\log\left(\frac{T_B \mu_{\rm m}^2 v^4_{\rm rel}}{4 \pi Q^2 \alpha_{\rm EM}^3 n_e}\right),$$
 where $Q$ is the millicharge, $\mu_{\rm m}$ is the DM-baryon reduced mass, $v_{\rm rel}$ is the DM-baryon relative velocity and $n_e\equiv x_e n_H$.  A further approximation $\mu_{\rm m} v_{\rm rel}^2 \to 3 T_B$~\cite{Liu:2019knx} leads to Eq.~(\ref{sigma0}). However, this approximation leaves out two effects:  first the DM temperature is different from the baryon temperature, and second it does not include the bulk relative velocity. An improved
  analysis should take both of these into account. The numerical size of these corrections is yet unknown. In the analysis here, we use Eq.~(\ref{sigma0}). We thank Hongwan Liu for communications and for pointing out the typos in Eq.~(10) of~\cite{Liu:2019knx} and the connection between Eq.~(\ref{sigma0}) and Eq.~(10) of their paper and the caveats regarding the approximation mentioned
 above.}

The relative velocity $V_{DB}$ between DM and baryons can produce a drag force $D(V_{DB})$ defined as $D(V_{DB})\equiv \text{d}V_{DB}/\text{d}t$~\cite{Munoz:2015bca}. The full expression for $D(V_{DB})$ is 
\begin{equation}
    D(V_{DB})=\sum_{t=e,p}\sigma_0\frac{f_{\rm dm} \rho_D+\rho_B}{m_t+m_D}\frac{\rho_t}{\rho_B}\frac{1}{V_{DB}^2}F(r_t).
    \label{eqn:drag}
\end{equation}
Here $f_{\rm dm}$ is the fraction of  millicharged DM, $\rho_B=\Omega_B(1+z)^3\rho_{\rm crit}$ is the energy density of baryons,  and $\rho_D=\Omega_c(1+z)^3\rho_{\rm crit}$ is the energy density of cold dark matter (CDM).
In the analysis
we take $\Omega_B= 0.0456$, $\Omega_c=0.227$, and $\rho_{\rm crit}=3H_0^3/(8\pi G_N)$, where $H_0=67.8$ km/s/Mpc (the Hubble   
parameter today), $G_N=6.67\times 10^{-11} \text{N kg}^{-2}\text{m}^2$ (the gravitational constant) and $\rho_t=x_e m_t n_H$.
The function $F(r_t)$ in Eq.~(\ref{eqn:drag}) is given by
\begin{equation}
    F(r_t)\equiv \mathrm{Erf}\left(\frac{r_t}{\sqrt{2}}\right)-\sqrt{\frac{2}{\pi}}\,r_t\,e^{-r_t^2/2}.
\end{equation}
Here $r_t\equiv V_{DB}/\bar{u}_t$,  where $\bar{u}_t$ is the average velocity due the thermal motion defined by
\begin{equation}
\bar{u}_t=\sqrt{T_B/m_t+T_D/m_D}\,.
\end{equation}
The EDGES signal is more prominent during the epoch where DM and baryons are tightly coupled together which means that the target particle includes both electrons and protons.
The coupling between baryons and DM induces
 heat transfer between the baryon fluid and DM which affects the evolution of the dark matter and baryon temperatures.
 The quantities that couple the  evolution  of $T_B$ and $T_D$ are~\cite{Munoz:2018pzp}
\begin{align}
\label{QB}
    \Dot{Q}_B&={f_{\rm dm}\frac{\rho_D}{m_D}}\frac{x_e}{1+f_{\rm He}}\sum_{t=e,p}\frac{m_D m_t}{(m_D+m_t)^2}\frac{\sigma_0}{\bar{u}_t}\left[\sqrt{\frac{2}{\pi}}\frac{e^{-r_t^2/2}}{\bar{u}^2_t}(T_D-T_B)+m_D\frac{F(r_t)}{r_t}\right], \\
\Dot{Q}_D&=n_{\rm H} x_e\sum_{t=e,p}\frac{m_D m_t}{(m_D+m_t)^2}\frac{\sigma_0}{\bar{u}_t}\left[\sqrt{\frac{2}{\pi}}\frac{e^{-r_t^2/2}}{\bar{u}^2_t}(T_B-T_D)+m_t\frac{F(r_t)}{r_t}\right],
\label{QD}
\end{align}
{where $f_{\rm He}\equiv n_{\rm He}/n_{\rm H}\approx 0.08$ is the ratio of the helium to hydrogen number densities.}

The evolution equations of baryon and DM temperatures, $T_B$ and $T_D$, of the relative velocity $V_{DB}$ and of the ionization rate $x_e$
are given by~\cite{Munoz:2018pzp,Kovetz:2018zan}
\begin{align}
\label{td}
(1+z)\frac{\text{d}T_D}{\text{d}z}&=2T_D {+\frac{\Gamma_\phi}{H(z)}(T_D-T_\phi)}-\frac{2}{3H(z)}\dot{Q}_D, \\
\label{tb}
(1+z)\frac{\text{d}T_B}{\text{d}z}&=2T_B+\frac{\Gamma_c}{H(z)}(T_B-T_\gamma)-\frac{2}{3H(z)}\dot{Q}_B, \\
\label{xe}
H(z)(1+z)\frac{\text{d}x_e}{\text{d}z}&=C\left[n_H\alpha_Bx_e^2-4(1-x_e)\beta_Be^{-3E_0/4T_{\gamma}}\right], \\
\label{vdb}
 (1+z)\frac{\text{d}V_{DB}}{\text{d}z}&=V_{DB}+\frac{D(V_{DB})}{H(z)},
\end{align}
where $H(z)$ is the Hubble parameter as a function of the redshift $z$,
\begin{equation}
    H(z)=H_0\sqrt{\Omega_m(1+z)^3+\Omega_\Lambda+\Omega_k(1+z)^2}.
\end{equation}
Here $\Omega_m=\Omega_B+\Omega_{c}$ is the matter energy density parameter (which includes dark matter, baryons and other non-baryonic matter of the visible sector), $\Omega_\Lambda=0.685$ is the dark energy density parameter and $\Omega_k=0.001$ is
 the curvature parameter~\cite{Planck:2018vyg}.
In Eq.~(\ref{tb}), $\Gamma_c$ represents the Compton interaction rate~\cite{Ali-Haimoud:2010tlj,Ma:1995ey}\footnote{We use natural units $k_B=\hbar=c=1$.}
\begin{equation}
\Gamma_c=\frac{64\pi^3\alpha^2T_\gamma^4}{135m_e^3}\frac{x_e}{1+x_e+f_{\rm He}},
\end{equation}
where $T_\gamma$ is the CMB photon temperature which evolves as $T_\gamma=2.726(1+z)$, $x_e$ is the free electron fraction which evolves according to Eq.~(\ref{xe}) and $m_e$ is the electron mass. Analogously, one can define the interaction rate $\Gamma_\phi$ which is proportional to $\lambda^2$. Since $\lambda\sim 10^{-3}$ or smaller,  this term in Eq.~(\ref{td}) makes a negligible contribution.
In Eq.~(\ref{xe}), $E_0=13.6$ eV is the ionization energy of hydrogen and $E_n\equiv E_0 n^{-2}$ is the $n^{\rm th}$  energy  level.
The quantity $C$ is the Peebles factor~\cite{Peebles:1968ja},
\begin{equation}
    C=\frac{\frac{3}{4}R_{\mathrm{Ly\alpha}}+\frac{1}{4}\Lambda_{2s,1s}}{\beta_B+\frac{3}{4}R_{\mathrm{Ly\alpha}}+\frac{1}{4}\Lambda_{2s,1s}},
\end{equation}
where $R_{\rm{Ly}\alpha}$ is the rate of escape of Ly$\alpha$ photons and is given by~\cite{AliHaimoud:2010dx}
\begin{equation}
    R_{\rm{Ly}\alpha}=\frac{8\pi H(z)}{3n_{\rm H} x_{1s}\lambda^3_{\rm{Ly}\alpha}},
\end{equation}
and $\Lambda_{2s,1s}=8.22\,\,\rm{s}^{-1}$ being the total $2s\to 1s$ two-photon decay rate.
 In the recombination epoch, the ground state population $x_{1s}\approx 1-x_e$. $\lambda_{\rm{Ly}\alpha}$ is the transition wavelength of a photon moving from the $n=2$ energy level to $1s$ state and can be calculated using $\lambda_{\rm{Ly}\alpha}=2\pi/E_{n1}$, where $E_{n1}=E_2-E_0=3/4 E_0$.
The quantity $\alpha_B$ in Eq.~(\ref{xe}) is the case-B recombination coefficient\footnote{The method used to estimate this coefficient is based on a simple assumption that the photons produced during recombination directly to the $n=1$ state are excluded since they are immediately reabsorbed by the gas. This is known as case-B recombination.} which is described by the fitting function~\cite{1991A&A...251..680P}
\begin{equation}
    \mathcal{\alpha}_B(T_B)=10^{-13}\frac{a(10^{-4}T_B)^b}{1+c(10^{-4}T_B)^d},
\end{equation}
where $a=4.309$, $b=-0.6166$, $c=0.6703$, and  $d=0.5300$. Further, $\beta_B$ in Eq.~(\ref{xe})
 is the corresponding photoionization rate which can be obtained from $\alpha_B$ by the principle of detailed balance~\cite{AliHaimoud:2010dx}
\begin{equation}
    \beta_B(T_\gamma)=\frac{g_e}{4}e^{E_2/T_\gamma}n_H\alpha_B(T_B=T_\gamma),
\end{equation}
where $g_e$ is defined by
\begin{equation}
g_e=\left(\frac{\mu_eT_\gamma}{2\pi}\right)^{3/2}\frac{1}{n_{\rm{H}}},
\end{equation}
and $\mu_e=m_em_p/(m_e+m_p)$.

\section{The brightness temperature of the 21-cm line}\label{sec:T21}

The quantity of interest in explaining the EDGES result is the  brightness temperature $T_{21}$
of the 21-cm line defined by~\cite{Kovetz:2018zan}
\begin{equation}
    T_{21}(z)=\frac{T_s-T_{\gamma}}{1+z}(1-e^{-\tau}),
    \label{eqn:T21}
\end{equation}
where $\tau$ is the optical depth for the transition given by
\begin{equation}
    \tau=\frac{3T_*A_{10}\lambda_{21}^3
    n_{\rm HI}}{32\pi T_sH(z)}.
\end{equation}
Here $n_{\rm HI}$ is given by
\begin{equation}
n_{\rm HI}=n_{\rm H}(1-x_e),
\end{equation}
where $A_{10}=2.869\times 10^{-15}\,\text{s}^{-1}$ is
the Einstein coefficient of the hyperfine transition, $T_*$ = 0.068 K, the energy difference between the two hyperfine levels
and $\lambda_{21}=21.1$ cm, the transition wavelength.  We assume full Ly$\alpha$ coupling
(i.e. $T_s=T_B$) to obtain $T_{21}$~\cite{Madau:1996cs}. Note that $T_{21}$ as defined in Eq.~(\ref{t21z}) can be obtained starting from Eq.~(\ref{eqn:T21}). See Refs.~\cite{Pritchard:2011xb,McQuinn:2005hk} for a derivation.

The sky-averaged 21-cm temperature is~\cite{Pritchard:2010pa,Cohen:2017xpx}
\begin{equation}
    \overline{T}_{21}\equiv\int \text{d}V_{DB}\mathcal{P}(V_{DB})T_{21}[T_B(V_{DB})].
  \label{T21av}
\end{equation}
The initial velocity probability distribution function is
\begin{equation}
 \mathcal{P}(v)=\frac{4\pi v^2e^{-3v^2/(2v_{\mathrm{rms}}^2)}}{(2\pi v_{\mathrm{rms}}^2/3)^{3/2}},
\end{equation}
where the root-mean-square velocity $v_{\mathrm{rms}}=29$ km/s is taken from Ref.~\cite{Munoz:2018pzp}. For a more detailed discussion on $v_{\mathrm{rms}}$, the reader
is referred to Ref.~\cite{Boddy:2018wzy}. The onset of drag between the dark matter and baryonic fluids could have been considerable much earlier in the cosmological history causing a much smaller $v_{\mathrm{rms}}$ by the time $z\sim 1100$. We have checked the effect of this on $\overline{T}_{21}$ by varying $v_{\mathrm{rms}}$ from 29 km/s to 1 km/s. We notice, for example, a $\sim 1.2\%$ change in $\overline{T}_{21}$ for benchmark (c), i.e., from $\sim -495$ mK to $\sim -489$ mK and up to $4\%$ for other benchmarks.

\section{Numerical analysis: millicharged DM relic density, temperature evolution and fit to EDGES data}\label{sec:numerical}

The general assumption made when solving the baryon and dark matter temperature evolution equations, i.e., Eqs.~(\ref{td})$-$(\ref{vdb}), is that $T_D=0$ at $z\sim 1000$. In this section we show that this comes about naturally from a particle physics model after taking into account all possible scattering processes that DM undergoes. 

In our model, the dark sector which contains the dark fermions, dark photons and the massless scalar is heated differently than the visible sector by the inflaton in the early universe. Thus the Boltzmann equations, Eqs.~(\ref{y1})$-$(\ref{yphi}), depend on the evolution of the temperature ratio $\eta$ given by Eq.~(\ref{y3}).
To take account of the assumption that the hidden sector is initially cooler than the visible sector, the coupled equations, Eqs.~(\ref{y1})$-$(\ref{y3}), are solved numerically with the initial conditions $\eta_0=1000$ and {$Y_D^0=Y^0_{\gamma'}=Y^0_{\phi}=0$}. The hidden sector is then slowly heated by energy injection from the visible sector owing to the weak coupling between the two sectors. The energy exchange between the sectors is represented by the source term, Eq.~(\ref{y4}), and by the different thermally averaged cross sections of Eqs.~(\ref{y1})$-$(\ref{y3}). Thermal averaging can only be used if the particle species have formed a thermal bath at some point which is always true for the visible sector but it may not be the case for the dark sector. We will come back to this point later.

As noted in earlier works, the constraints derived from analyzing the Planck 2015 CMB data~\cite{Kovetz:2018zan} imply that the fraction of millicharged dark matter relevant to resolving the  EDGES effect cannot be more than $f_{\rm dm}\sim 0.4\%$ which greatly
reduces the allowed  parameter space of the model. There is also a mass-dependent lower bound below which a DM fraction can no longer be distinguished from the baryonic fluid~\cite{Kovetz:2018zan,Boddy:2018wzy}. Further constraints on millicharged DM come from SLAC~\cite{Prinz:1998ua}, SN1987A~\cite{Chang:2018rso} and CMB (Planck 2015 data)~\cite{Boddy:2018wzy}. A dark photon which couples via mass mixing to the visible sector is also constrained by several experiments such as the electron and muon $g-2$~\cite{Endo:2012hp}, BaBar~\cite{Lees:2014xha}, {CHARM~\cite{Bergsma:1985qz,Tsai:2019mtm,Gninenko:2012eq}}, NA48~\cite{Batley:2015lha}, E137~\cite{Andreas:2012mt,Bjorken:2009mm}, NA64~\cite{Banerjee:2018vgk,Banerjee:2019hmi}, E141~\cite{Riordan:1987aw}, $\nu$-CAL~\cite{Blumlein:1990ay,Blumlein:1991xh,Tsai:2019mtm} {and LHCb~\cite{LHCb:2017trq,LHCb:2019vmc}}.
In Table~\ref{tab1} we display three representative benchmarks which satisfy all constraints and can explain the EDGES result. A DM fraction of 0.3\% makes any constraints from DM direct detection experiments very weak as well as those from indirect detection in the dominant {$D\bar{D}\to\gamma'\gamma'\to 4e$}	 channel~\cite{Ade:2015xua}. Note that in Table~\ref{tab1}, $\epsilon=M_2/M_1$ and since the mixing is small one has $m_{\gamma'}\sim M_1$.  We note that in the analysis of Table~\ref{tab1}, we set the kinetic mixing parameter $\delta=0$.
Inclusion of $\delta$ affects the analysis only to order $\delta^3$ and since kinetic mixing is typically in the range $10^{-10}-10^{-9}$,
 it makes negligible contribution to the analysis.

\begin{table}[H]
\caption{\label{tab1}
The benchmarks used in this analysis where the  fraction of the relic density is $f_{\rm dm}=\Omega h^2/(\Omega h^2)_{\rm Planck}\sim 0.3\%$, $\delta=0$ {and $\lambda=10^{-5}$}. All masses are in MeV.}
\begin{center}
\begin{tabular}{|cccccccc|}
\hline\hline
Model & $m_D$ & $m_{\gamma'}$ & $\epsilon_D~(\times 10^{-5})$ & $\epsilon~(\times 10^{-5})$ & $g_X$ & $Q_X$ & $\Omega h^2~(\times 10^{-4})$ \\
\hline
(a) & 16 & {46.0} & 1.94 & 3.50 & 0.20 & 1.0 & {3.38} \\
(b) & 12 & {34.2} & 2.22 & 8.00 & 0.10 & 1.0 & {3.49} \\
(c) & 7 & {20.5} & 1.86  & 4.00 & 0.25 & 2/3 & {3.41} \\
\hline\hline
\end{tabular}
\end{center}
\label{tab1}
\end{table}

Even though the couplings between the dark sector and SM particles is weak, the coupling among the dark sector particles is large enough to establish a thermal bath. Chemical equilibrium is established when the interaction rate of number changing processes, such as $D\bar{D}\to\gamma'\gamma'$, becomes larger than the Hubble parameter. Even after the particle species fall out of chemical equilibrium, kinetic equilibrium can still be maintained in the dark sector via scattering with massless degrees of freedom, i.e. $D\phi\to D\phi$. We show in the left panel of Fig.~\ref{fig2} the yield of {$D$, $\gamma'$ and $\phi$} as a function of the hidden sector temperature $T_h$ for benchmark (b). The yield starts increasing at high temperatures as the SM injects energy into the hidden sector. The rise in the yield of $\phi$ is delayed since the scalar particle is produced only through interactions with $D$ as indicated in Eq.~(\ref{yphi}). Shortly after, the dark sector particles {$D$ and $\gamma'$} enter in chemical equilibrium as indicated in the figure {while $\phi$ joins the thermal bath later depending on the size of $\lambda$}. After that, the particle yields start tracing their equilibrium distribution shown by the green dashed line. The yields of {$D$, $\gamma'$ and $\phi$} reach a maximum and plateau for a period of time. Annihilations of $D\bar{D}$ to $\gamma'\gamma'$ and to SM particles deplete the DM relic density before it eventually departs from its equilibrium number density and freezes out at $T_h\sim 0.4$ MeV. This corresponds to $x=m_D/T_h\sim 30$, a typical value around which freeze-out occurs in standard scenarios. Thus the freeze-out mechanism completely determines the DM relic density as it erases all memory pertaining to the specifics of the production mode prior to equilibrium. {The yield of $\phi$ increases slightly above its equilibrium value due to processes such as $D\bar{D}\to\phi\phi, \phi\gamma, \phi\gamma'$ after which it remains constant following freeze-out. The yield of $\phi$ is shown for two values of $\lambda$ (solid and dotted black lines) where a value three orders of magnitude higher results in a larger yield but has a moderate effect on the relic density where we observe a decrease of less than 5\%.} The dark photon's yield shows a dramatic drop as it is eliminated from the bath via the decays $\gamma'\to i~\bar{i},D\bar{D}$. The cyan dashed line shows the equilibrium yield of $D$ as a function of the visible sector temperature, $Y^{\rm eq}_D(T)$. Notice that $Y_D$ and $Y_{\gamma'}$ come close to $Y^{\rm eq}_D(T)$ and so the hidden sector and the visible sector never fully thermalize. This can be seen in the right panel of Fig.~\ref{fig2} where we plot $\xi\equiv 1/\eta=T_h/T$ as a function of the visible temperature $T$. It is clear that the hidden sector is heated by the visible sector and reaches a maximum of $\xi\sim 0.8$ followed by some oscillations before leveling off at $\xi=0.5$. Let us explain the origin of the oscillations in $\xi$ which we divide into three regions I, II and III.

\begin{figure}[H]
    \centering
  \includegraphics[width=0.49\textwidth]{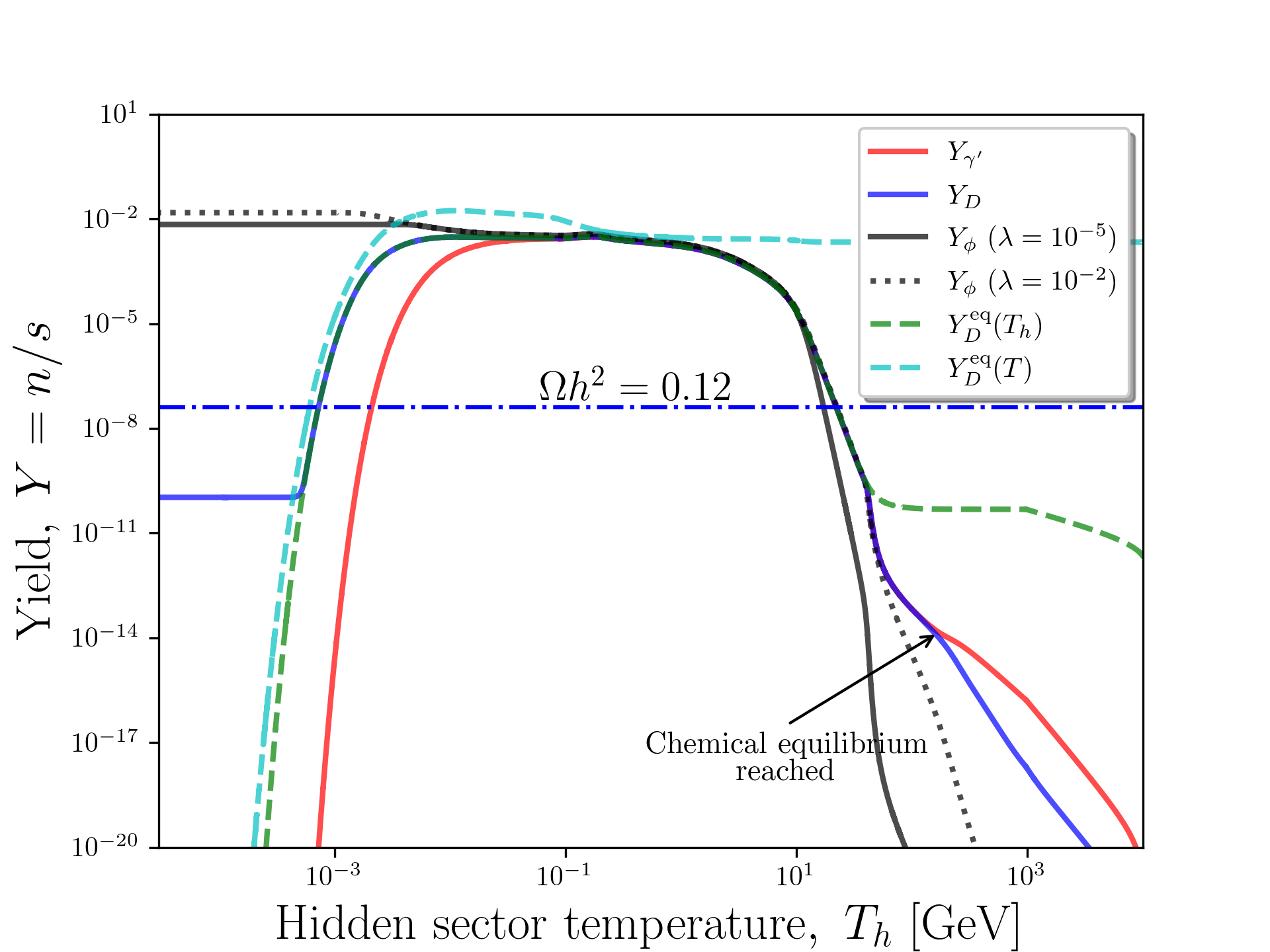}
  \includegraphics[width=0.49\textwidth]{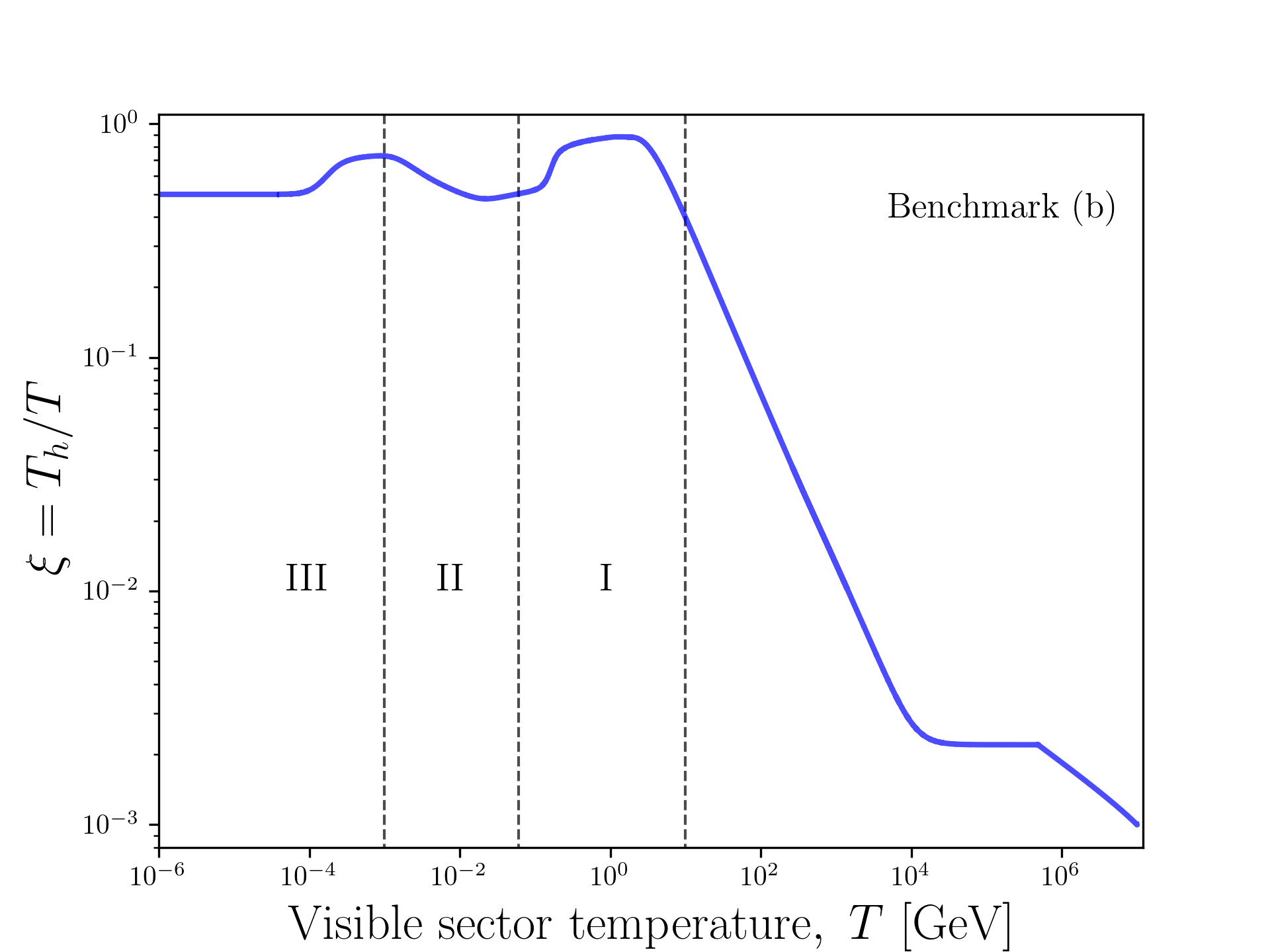}
    \caption{Left panel: {the yields of the dark fermion, dark photon and the scalar $\phi$} as a function of the hidden sector temperature $T_h$ for benchmark (b). Right panel: evolution of $\xi$ as  a function of the visible sector temperature $T$ for benchmark (b). The ratio levels off at $\xi= 0.5$.}
    \label{fig2}
\end{figure}

As the temperature drops, the source term $j_h$ in Eq.~(\ref{y4}) becomes subdominant in comparison to the energy density and can be dropped. In this case, Eq.~(\ref{y3}) can be further simplified and written in terms of $\xi$ as
\begin{equation}
\frac{d\ln\xi}{d\ln T_h}=1-\frac{1+\frac{1}{4}\frac{d\ln g^h_{\rm eff}}{d\ln T_h}}{1+\frac{1}{4}\frac{d\ln g^v_{\rm eff}}{d\ln T}}.
\label{deta}
\end{equation}
Thus, the temperature evolution after this point depends entirely on the available degrees of freedom and the manner in which they decouple as the universe cools down. We present in Fig.~\ref{fig3} the energy density effective degrees of freedom, $g_{\rm eff}$, (left panel) and $d\ln g_{\rm eff}/d\ln T$ (right panel) in the visible (blue line) and hidden (red line) sectors. The most dramatic changes in the visible sector degrees of freedom occur during the QCD phase transition, neutrino decoupling and electron-positron annihilation as shown by three different regions indicated in the left panel of Fig.~\ref{fig3} (they correspond to the regions I, II and III in the right panel of Fig.~\ref{fig2}). The situation is simpler for the hidden sector where we show the specific example of benchmark (c). The drop in $g_{\rm eff}$ occurs between 1 MeV and $\sim 40$ MeV which is reasonable given that $m_D=7$ MeV and $m_{\gamma'}=20$ MeV in this example. After $D$, $\gamma'$ decoupling, the massless scalar $\phi$ is now the only remaining degree of freedom. The right panel of Fig.~\ref{fig3} shows $d\ln g_{\rm eff}/d\ln T$ with the three main regions indicated as in the left panel. In region I, we notice a sharp rise in $d\ln g_{\rm eff}^v/d\ln T$ due to the QCD phase transition and so the second term on the right hand side of Eq.~(\ref{deta}) becomes small compared to one, therefore $d\ln\xi/d\ln T_h>0$ which causes the drop in $\xi$ to the left side of region I in Fig.~\ref{fig2} (right panel). Note that region I is not entirely dominated by the QCD phase transition since $j_h$ can still be efficient there and so Eq.~(\ref{deta}) is only valid towards the end of region I.

\begin{figure}[H]
    \centering
  \includegraphics[width=0.49\textwidth]{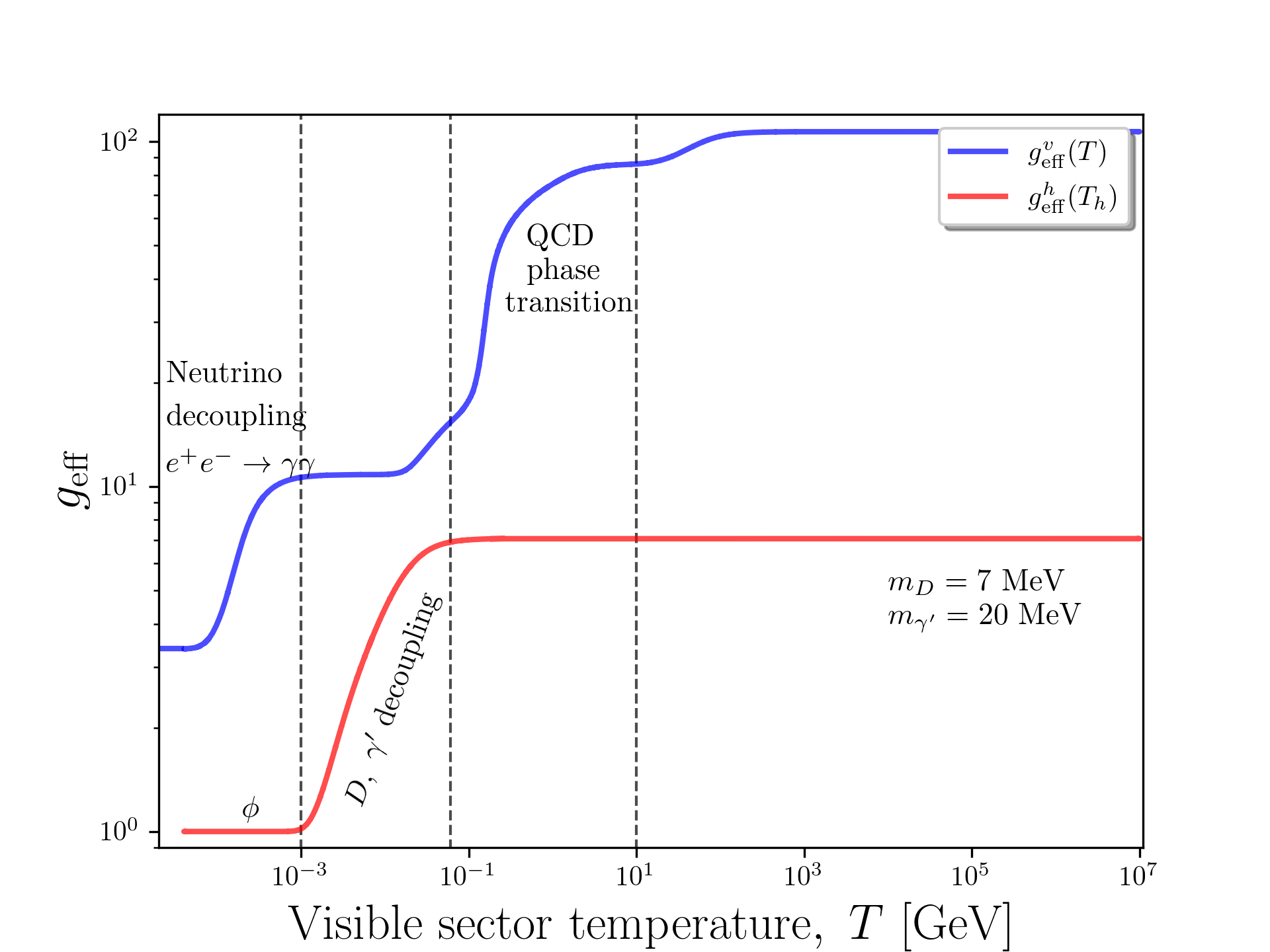}
  \includegraphics[width=0.49\textwidth]{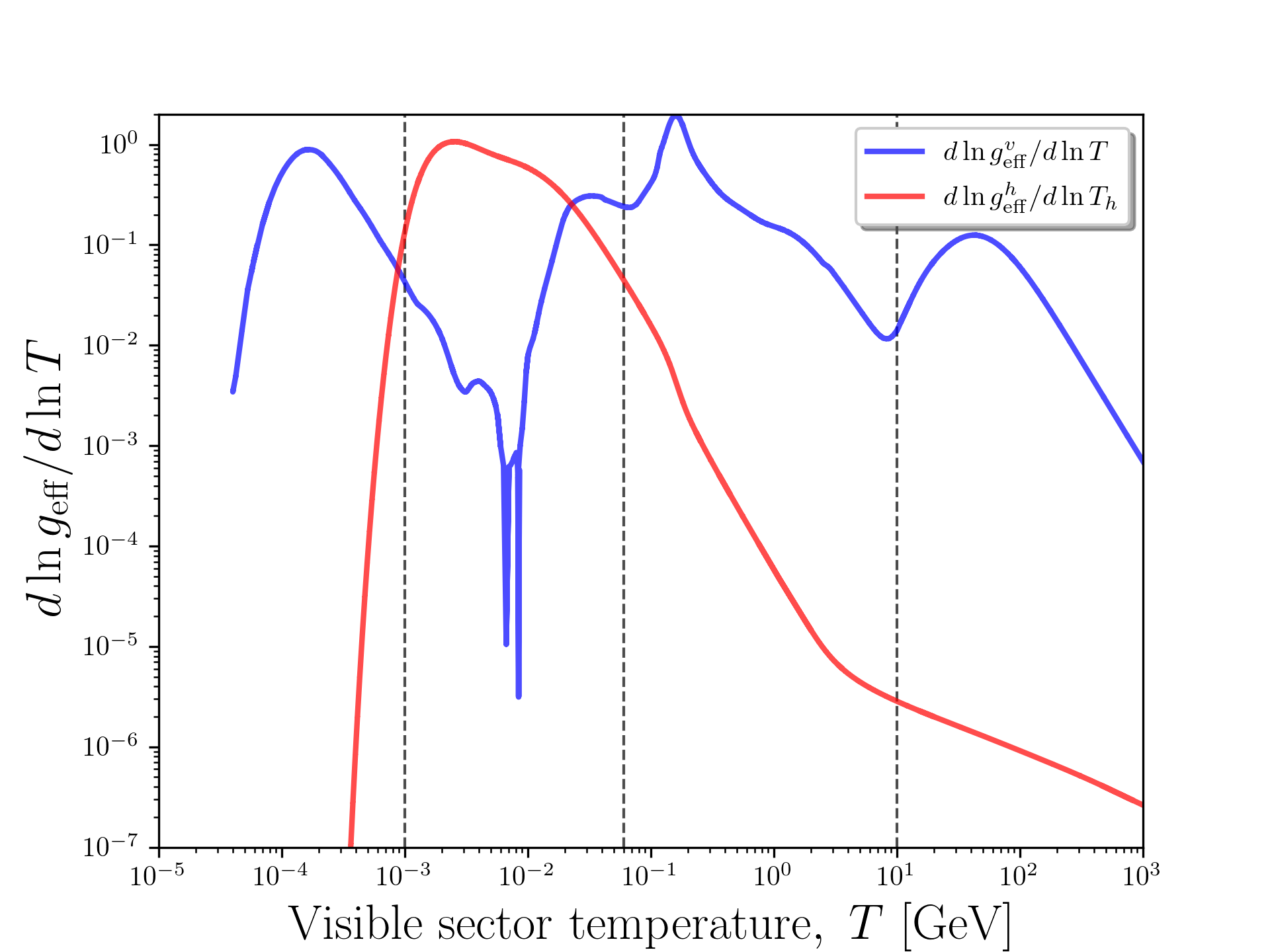}
    \caption{Left panel: a plot of $g_{\rm eff}$ as a function of the visible temperature $T$ for the visible (blue line) and hidden (red line) sectors. The different regions bounded by the dashed vertical lines represent regions with the largest changes in $g_{\rm eff}$. Those changes are exhibited in the right panel as $d\ln g_{\rm eff}/d\ln T$ where the same regions bounded by the vertical dashed lines are shown. The hidden sector degrees of freedom correspond to benchmark (c).}
    \label{fig3}
\end{figure}

Region II is dominated by the changes in $g_{\rm eff}$ of the hidden sector as one can see from the right panel of Fig.~\ref{fig3} that $d\ln g_{\rm eff}^h/d\ln T>d\ln g_{\rm eff}^v/d\ln T$. In this case $d\ln\xi/d\ln T_h$ turn negative which explain the rise in $\xi$ in region II. Finally, in region III we see a similar behavior as in region I since the visible degrees of freedom undergo a sudden change due to neutrino decoupling followed by electron-positron annihilation. This causes $d\ln\xi/d\ln T_h$ to turn positive resulting in a drop in $\xi$. Eventually $\xi$ levels off since $d\ln g_{\rm eff}^h/d\ln T=d\ln g_{\rm eff}^v/d\ln T=0$, hence $d\ln\xi/d\ln T_h=0$.

One needs to make sure that the remaining relativistic degree of freedom in the hidden sector due to $\phi$ does not contribute
 excessively to $\Delta N_{\rm eff}$. Thus one can write $\Delta N_{\rm eff}$ in terms of $\xi$ such that
\begin{equation}
\Delta N_{\rm eff}\simeq \frac{4\Delta n_b}{7}\left(\frac{11}{4}\right)^{4/3}\xi^4,
\end{equation}
where $\Delta n_b=1$ in our case of a massless real scalar in the hidden sector. From Fig.~\ref{fig2} (right panel), we have $\xi=0.5$ and so this gives us $\Delta N_{\rm eff}=0.138$ which is consistent with the experimental limits on $\Delta N_{\rm eff}\sim 0.2$.
Thus the extra relativistic degree of freedom can be easily diluted away owing to the temperature difference between the sectors.
We note that our  analysis of $\Delta N_{\rm eff}$ differs from previous analyses such as of Ref.~\cite{Boehm:2013jpa}
because it depends sensitively on the  ratio of the hidden sector temperature to the visible sector  temperature.

{This temperature difference is maintained through the processes $i~D\leftrightarrow i~D$ after chemical decoupling and remain active much later than BBN. As the universe cools, the dark fermions undergo kinetic decoupling which, using the recipe in Ref.~\cite{Bringmann:2009vf}, is illustrated in Fig.~\ref{fig3a} for benchmark (b). For small $x$, the temperature difference between $D$ (blue line) and the visible sector (green dashed line) is clear while $D$ is in the thermal bath created by $\phi$ in the hidden sector. Elastic scatterings between $D$ and the SM particles decouple at $x\sim 200$ and the two sectors appear to become in thermal contact. This is due to entropy redistribution among available degrees of freedom as the more massive ones decouple. Thermal equilibrium between the sectors remains until $D$ kinetically decouples from the dark sector thermal bath as $\phi D\leftrightarrow\phi D$ becomes weak at a temperature $\sim 1.2$ keV. Afterwards, $T_D$ drops as $1/a^2$, where $a$ is the scale factor. Note that the hidden sector no longer has a unified temperature $T_h$ and so after kinetic decoupling $T_h\neq T_D\neq T_\phi$.    }

\begin{figure}[H]
    \centering
  \includegraphics[width=0.65\textwidth]{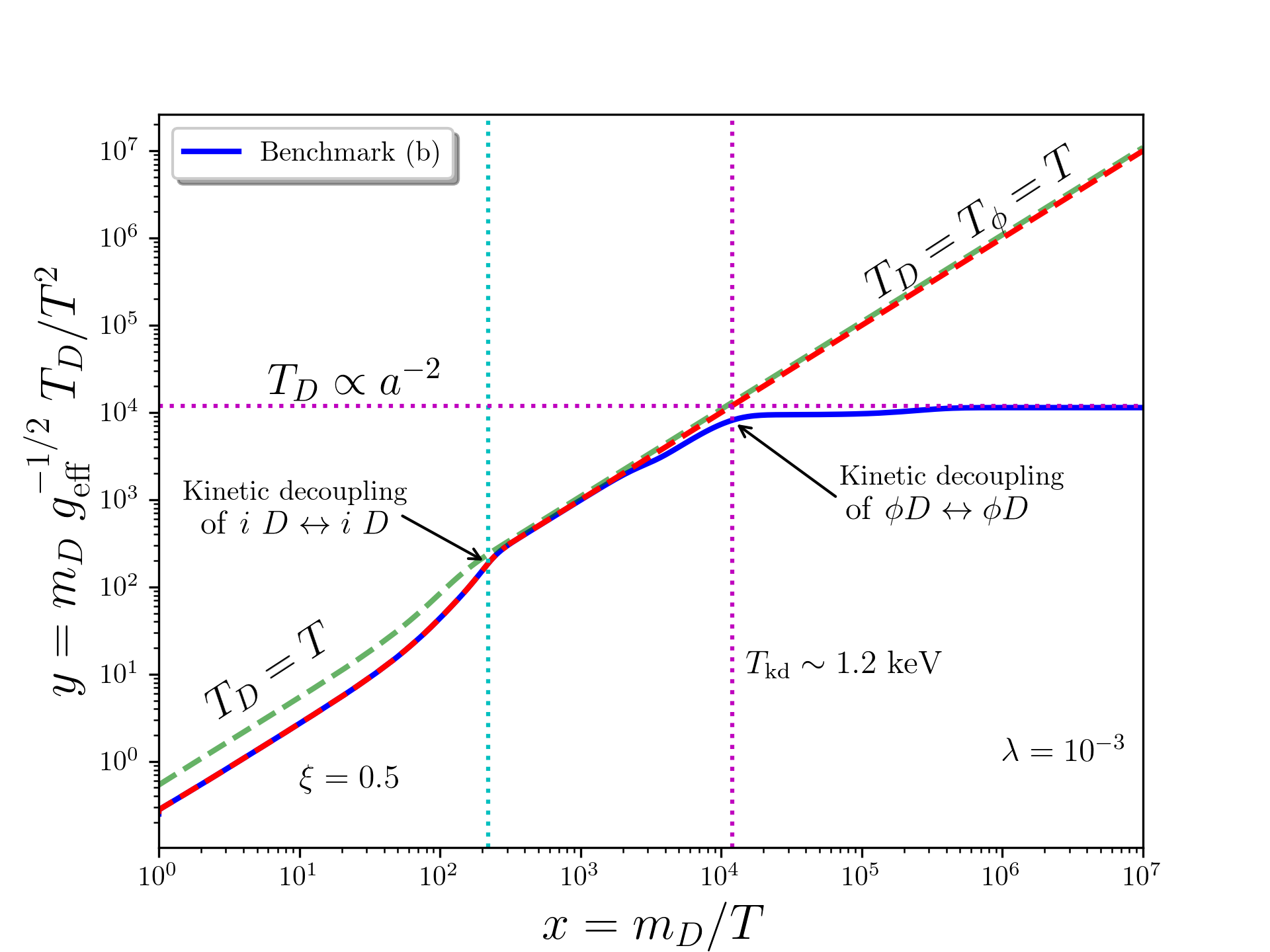}
    \caption{Evolution of the dark fermion temperature (blue line) as a function of $x=m_D/T$ showing the kinetic decoupling happening after the dark freeze-out. The green dashed line represents the visible sector temperature and the red dashed line is the hidden sector temperature.}
    \label{fig3a}
\end{figure}

Having established a temperature difference between the dark matter and the visible sector after solving Eqs.~(\ref{y1})$-$(\ref{y3}), we now turn our attention to the temperature evolution at much later times, i.e. at redshifts down to $z\lesssim 1700$. For this purpose, we solve the coupled equations, Eqs.~(\ref{td})$-$(\ref{vdb}), {with the initial condition $T_D\to 0$ and $T_B=T_{\rm CMB}$ at $z=1700$.} The results are shown in Fig.~\ref{fig4}.

\begin{figure}[H]
 \centering
 \includegraphics[width=0.32\textwidth]{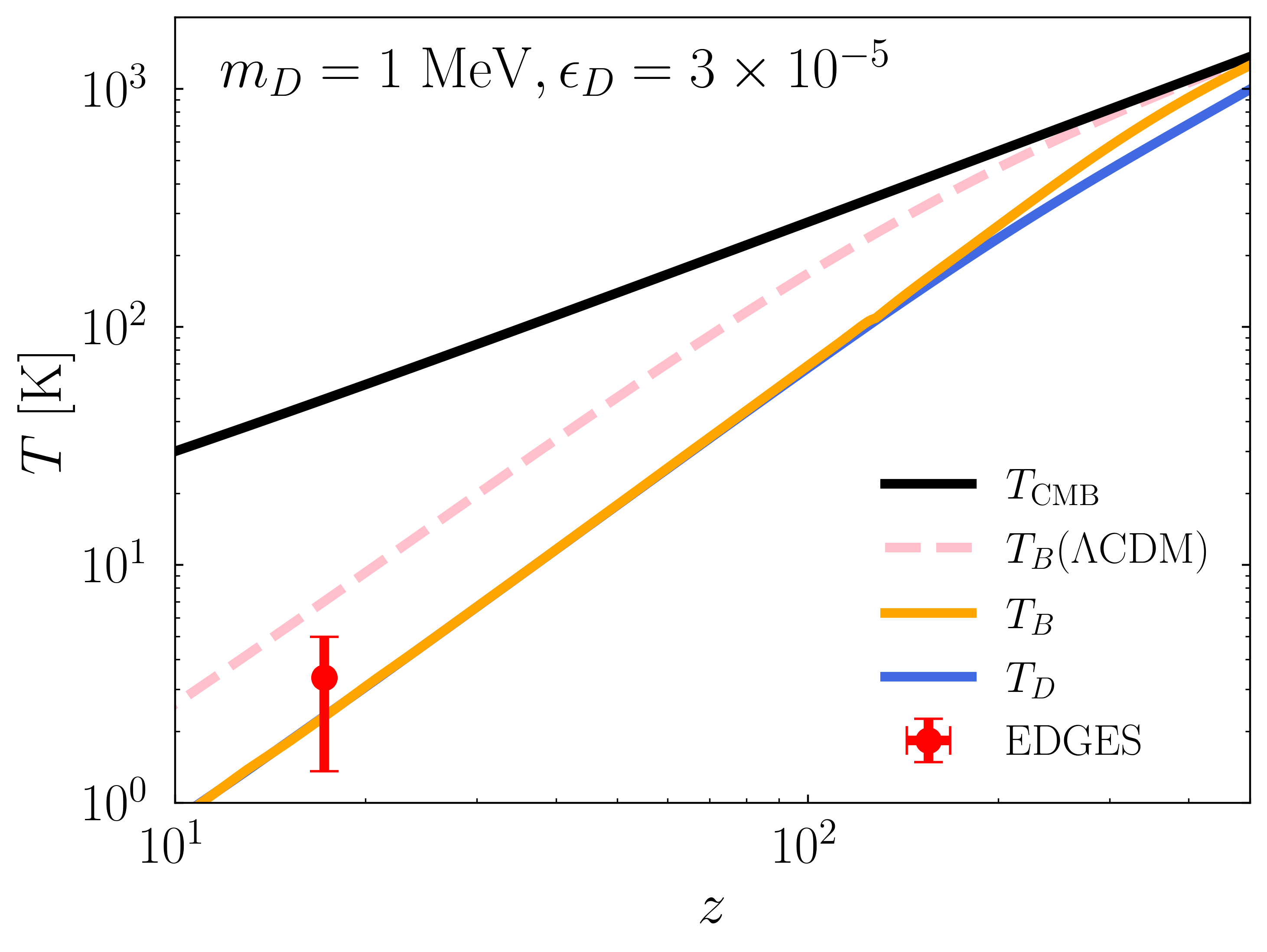}
  \includegraphics[width=0.32\textwidth]{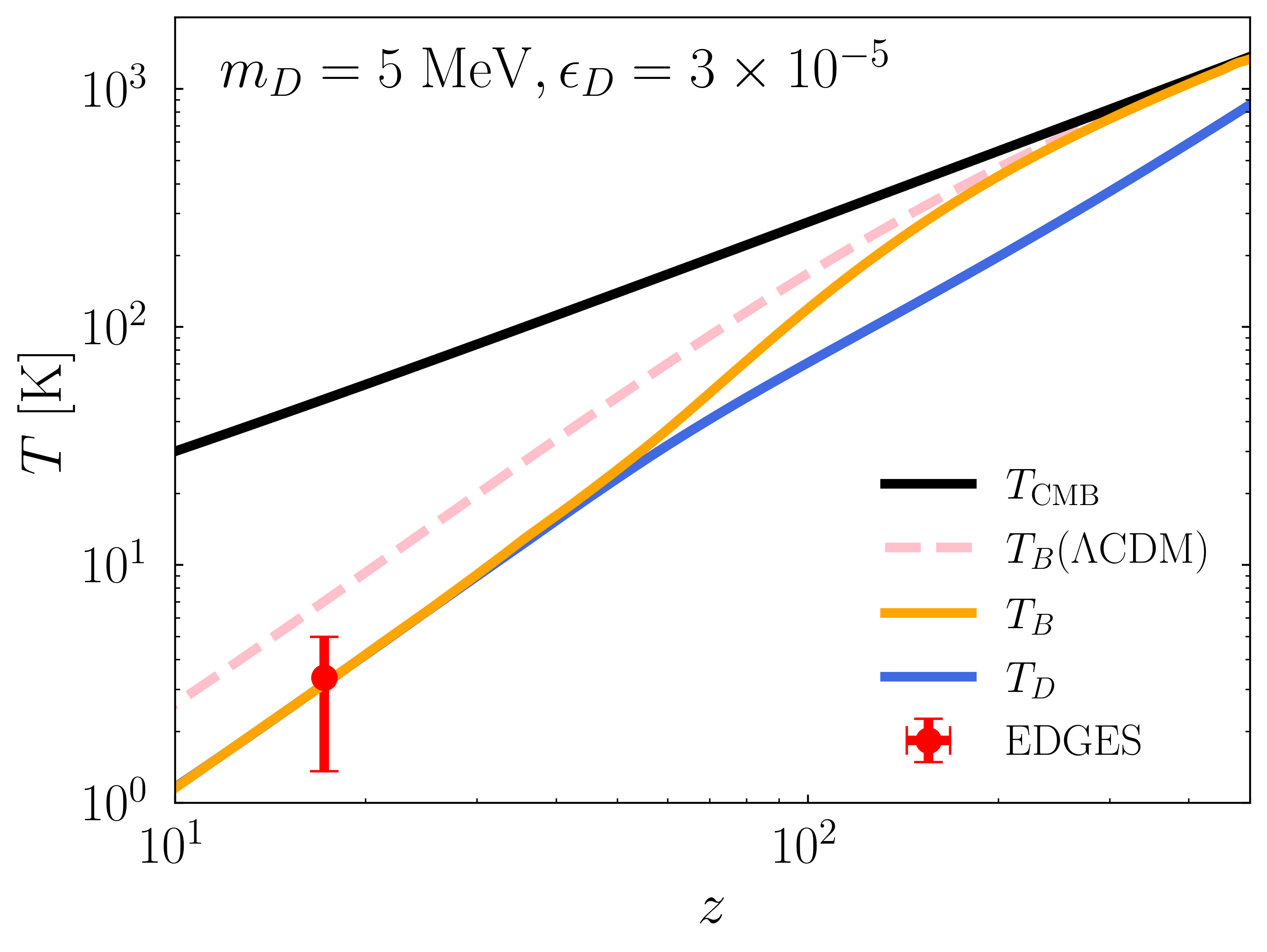}
  \includegraphics[width=0.32\textwidth]{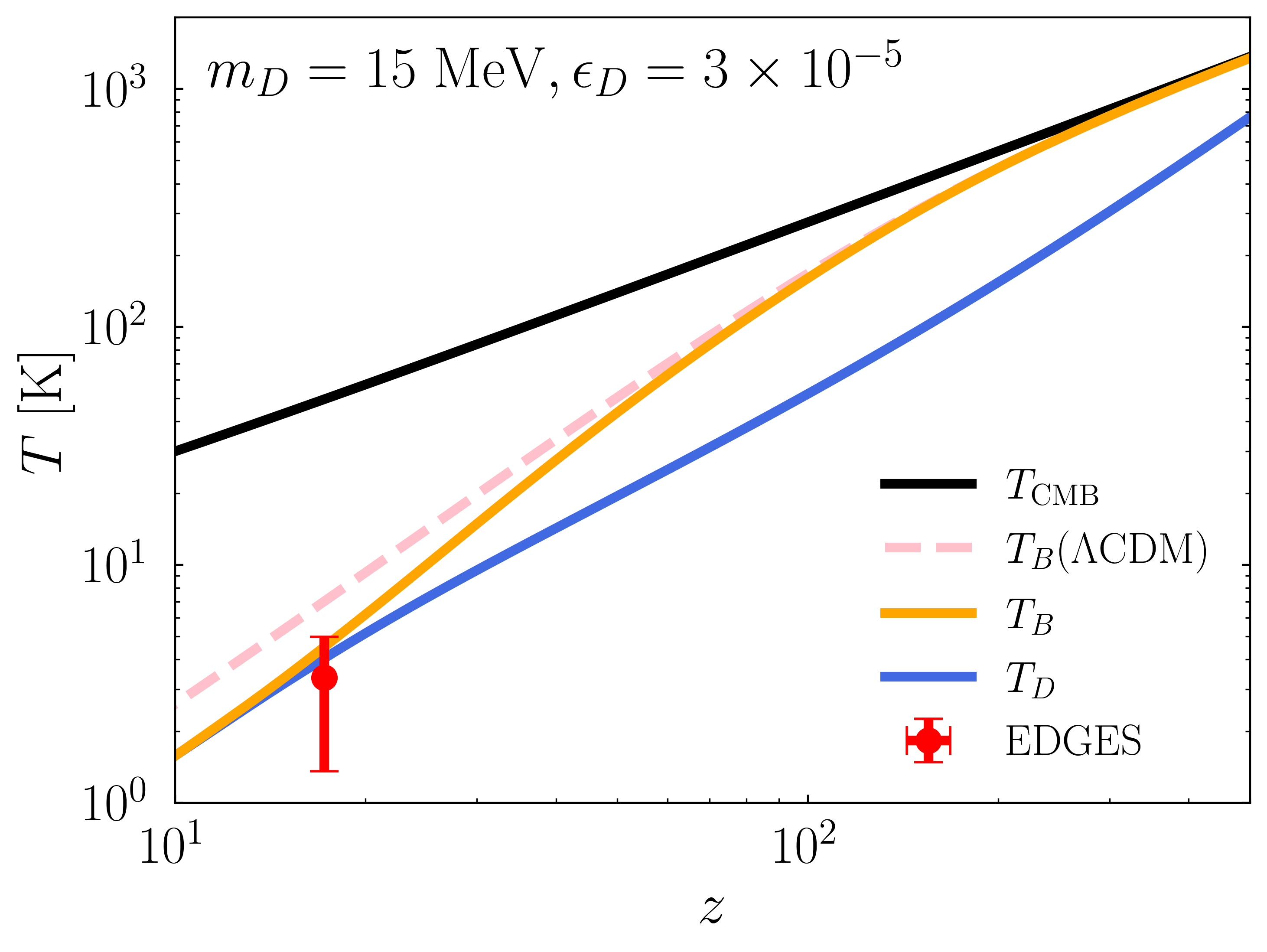}
 \caption{The temperature evolution of the CMB (black line), baryons (orange line), and DM (blue line) as a function of the redshift. The pink dashed line is the $T_B$ evolution in the $\Lambda \mathrm{CDM}$ model. The three panels correspond to three different values of $m_D$ and fixed $\epsilon_D=3\times 10^{-5}$ and $f_{\mathrm{dm}}=0.3\%$. It is seen that for $m_D$ small, baryons and DM have a faster heat exchange so they thermalize early on and DM cools baryons  more efficiently~\cite{Tashiro:2014tsa}. The red point with vertical error bars  in each of the three panels represents the EDGES measurement.}
    \label{fig4}
\end{figure}
The three panels Fig.~\ref{fig4}
show the temperature evolution of baryons and the DM for three values of $m_D$ while $\epsilon_D$ and $f_{\rm dm}$ are fixed
and consistent with the constraints of Fig.~\ref{fig6}. Notice that the CMB and the baryons were initially in thermal contact before they begin diverging as the universe cools. As the baryons fall out of equilibrium with the CMB, they begin to thermalize with the DM and move away from their predicted value in $\Lambda$CDM (dashed line) owing to the baryon-DM interaction which further cools down the baryons.

Next, we exhibit in Fig.~\ref{fig5}, the dependence of $\overline{T}_{21}$ on $\epsilon_D$ (left panel), on $f_{\rm dm}$ (middle panel) and on $m_D$ (right panel). The upper and the lower hatched regions are excluded by virtue of the value of $\overline{T}_{21}$ given by Eq.~(\ref{eqn:T21}).

\begin{figure}[H]
 \centering
 \includegraphics[width=0.32\textwidth]{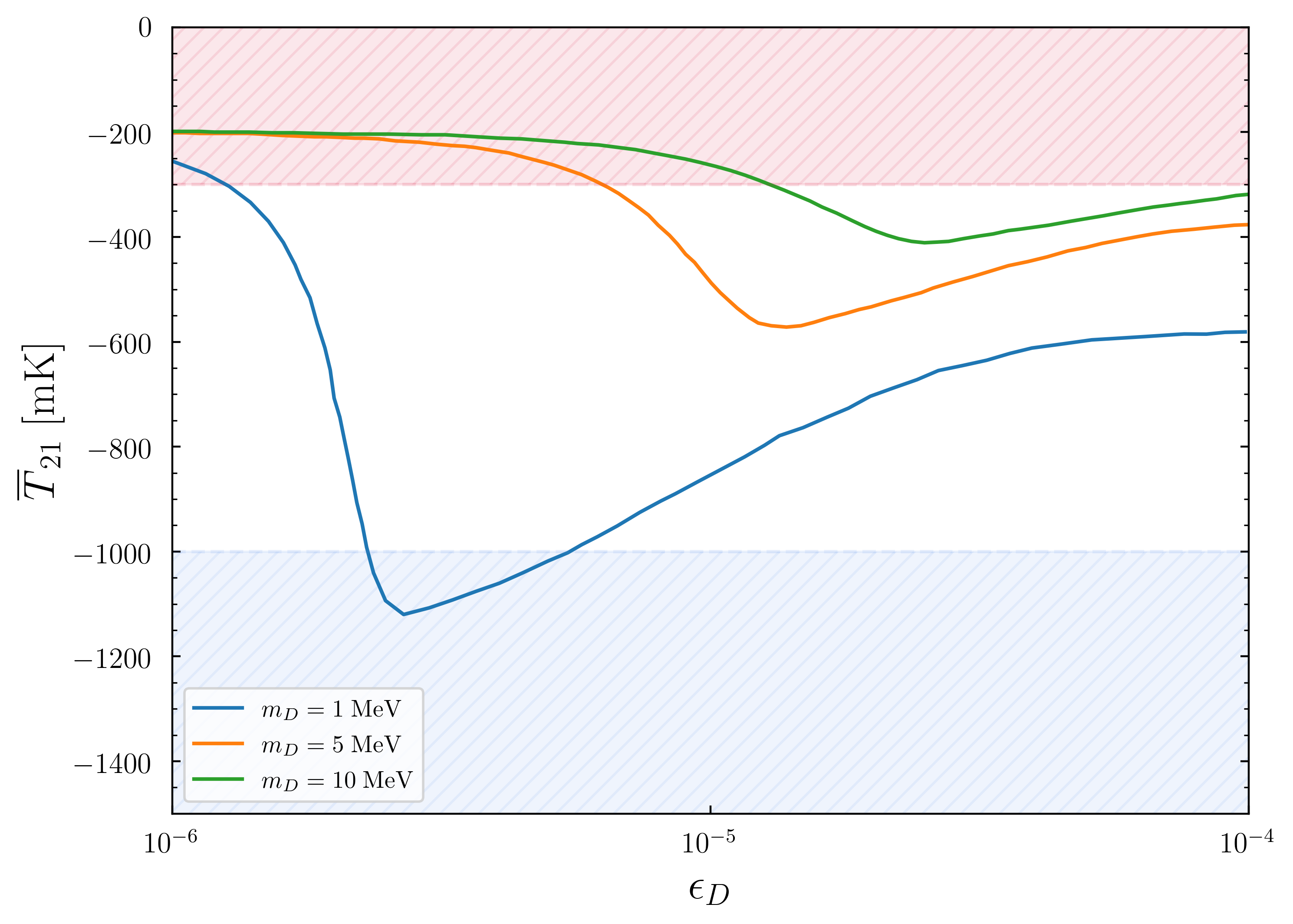}
  \includegraphics[width=0.32\textwidth]{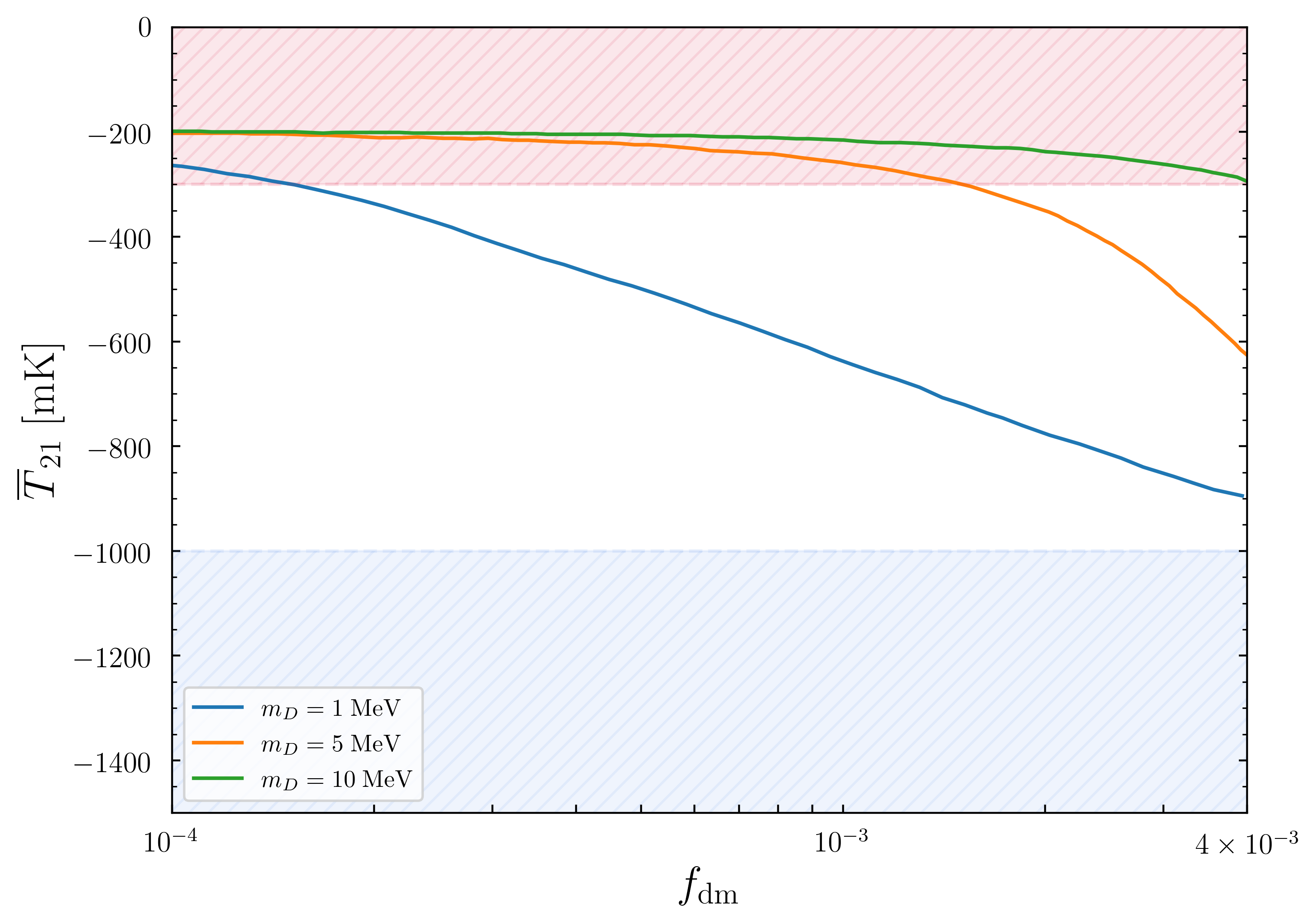}
  \includegraphics[width=0.32\textwidth]{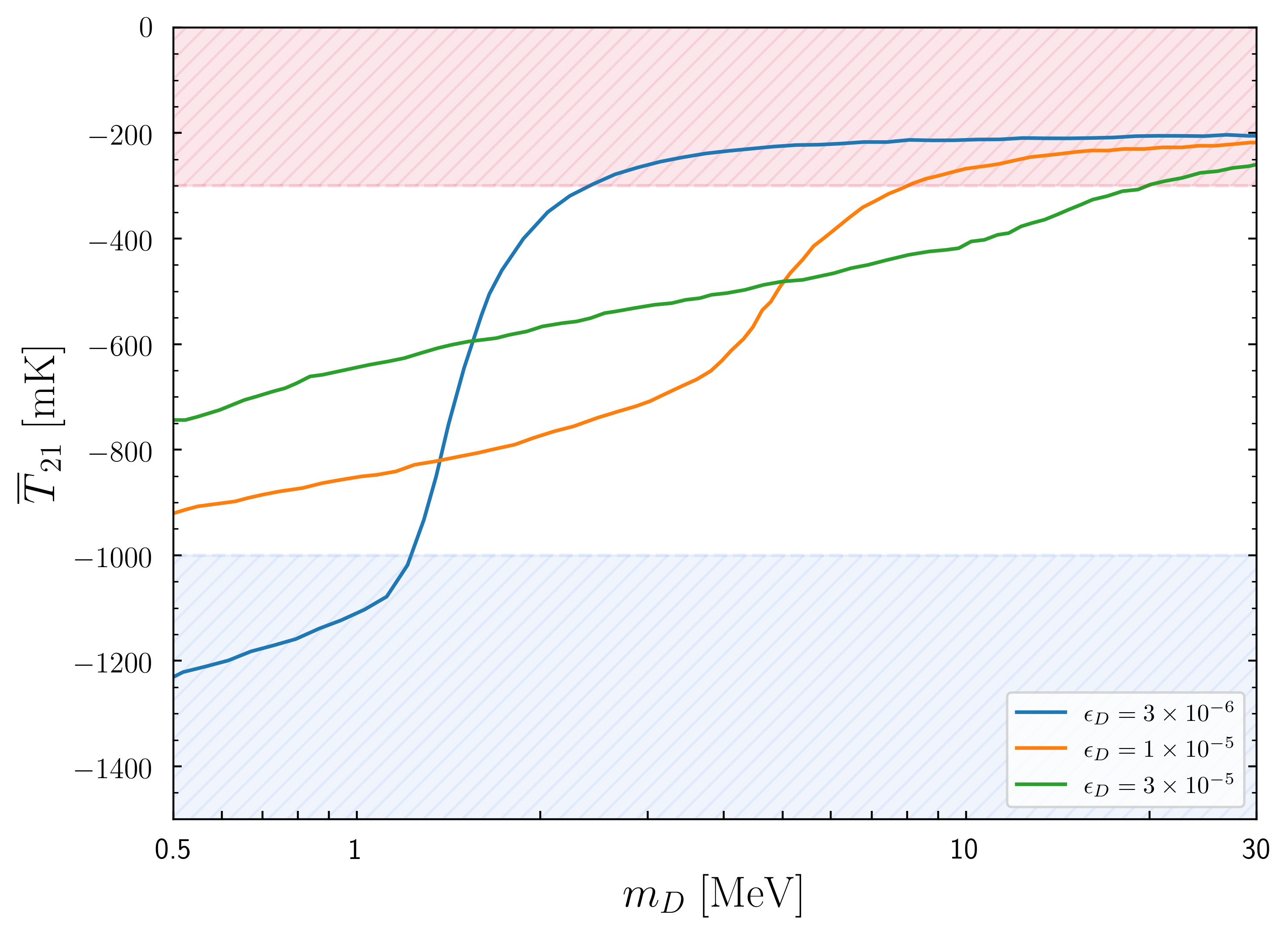}
 \caption{Exhibition of the dependence of $\overline{T}_{21}$ at $z=17.2$ on $\epsilon_D$ (left panel), on $f_{\rm dm}$ (middle panel) and on $m_D$ (right panel). Left panel: the three curves correspond to $m_D=1$ MeV (blue line),  $5\;\mathrm{MeV}$ (orange line) and $10\;\mathrm{MeV}$ (green line) with a fixed $f_{\mathrm{dm}}=0.3\%$. Middle panel: The three curves correspond to the same three values of $m_D$ as in the left
 panel except that here $\epsilon_D$ is fixed at $\epsilon_D=10^{-5}$. Right panel:  Here the three curves correspond to values of $\epsilon_D$ to be $3\times 10^{-6}$ (blue line),  $1\times 10^{-5}$ (orange line) and $3\times 10^{-5}$ (green line) with a fixed
 $f_{\mathrm{dm}}=0.3\%$.
  In all panels, the red hatched region is excluded by $\overline{T}_{21}>-300\;\mathrm{mK}$ while the blue hatched region is excluded by $\overline{T}_{21}<-1000\;\mathrm{mK}$. }
    \label{fig5}
\end{figure}

Finally in Fig.~\ref{fig6} we give the exclusion plots which display the region of the parameter of the Stueckelberg model
where the EDGES signal can reside consistent with all the constraints.
 The left panel of Fig.~\ref{fig6}   exhibits the relevant constraints in the $\epsilon_D$-$m_D$ plane on millicharged dark matter along with the allowed region consistent with EDGES result of
 $\overline{T}_{21}=-500^{+200}_{-500}$ mK.
 We also show contours for fixed $\overline{T}_{21}$ values in the allowed region. Note that the Planck constraint has been satisfied since we fixed $f_{\mathrm{dm}}=0.3\%$ consistent with $f_{\mathrm{dm}}\lesssim 0.4\%$~\cite{Kovetz:2018zan}. The right panel of Fig.~\ref{fig6} shows the constraints and the allowed regions in the $\epsilon$-$m_{\gamma'}$ plane. Note that here the plot is in the mass mixing $\epsilon$ and not the millicharge $\epsilon_D$. The two are related via Eq.~(\ref{milli-hid3}). The blue region corresponds to the parameter space giving $f_{\rm dm}\sim 0.3\%$ and consistent with the allowed DM millicharges in the left panel only within the region bounded by the hashed red line which corresponds to $\overline{T}_{21}>-300$ mK. Here we find that $m_{\gamma'}\sim 3m_D$ is able to deplete the DM relic density to the required fraction.

\begin{figure}[H]
 \centering
 \includegraphics[width=0.45\textwidth]{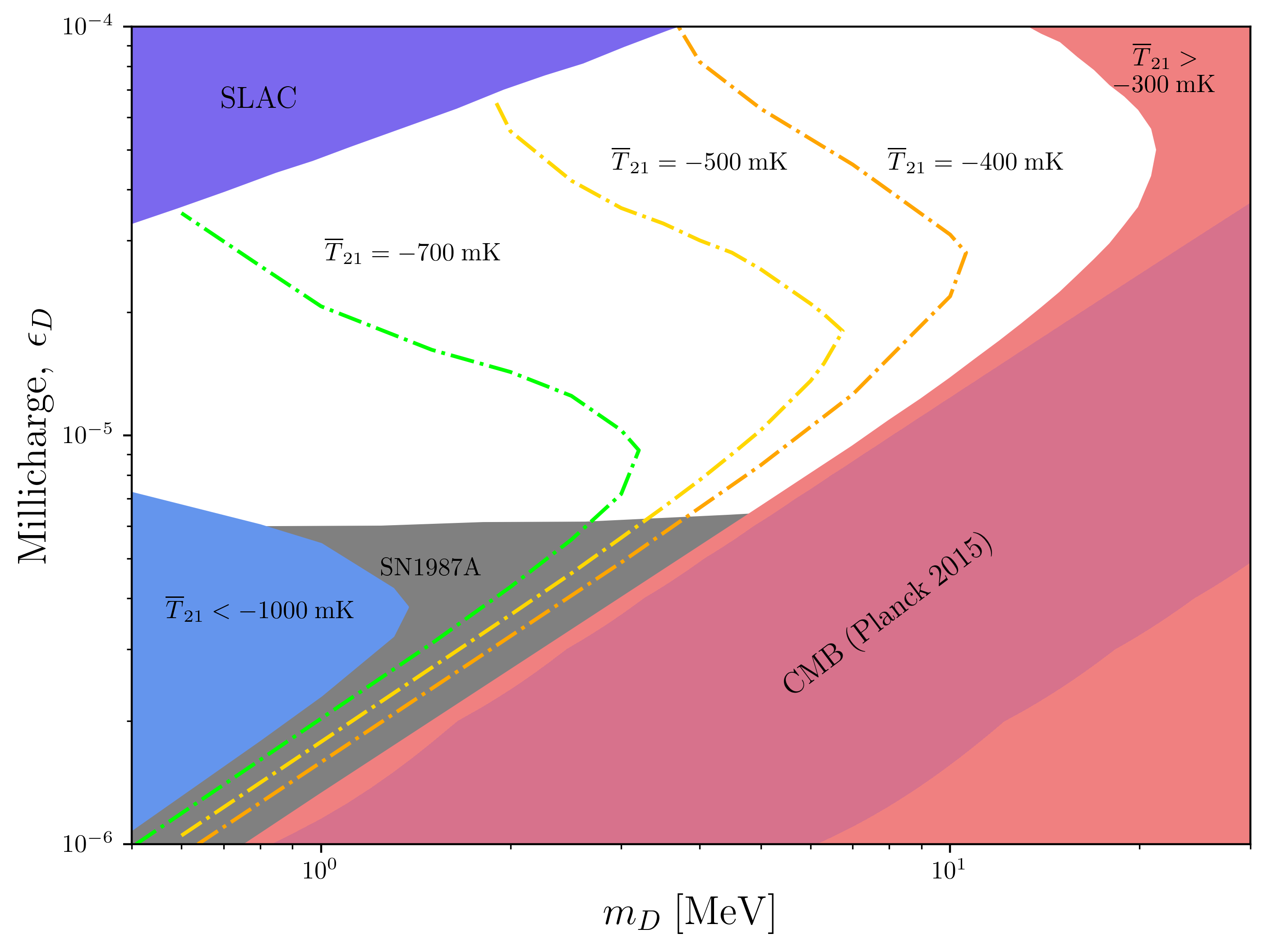}
 \includegraphics[width=0.49\textwidth]{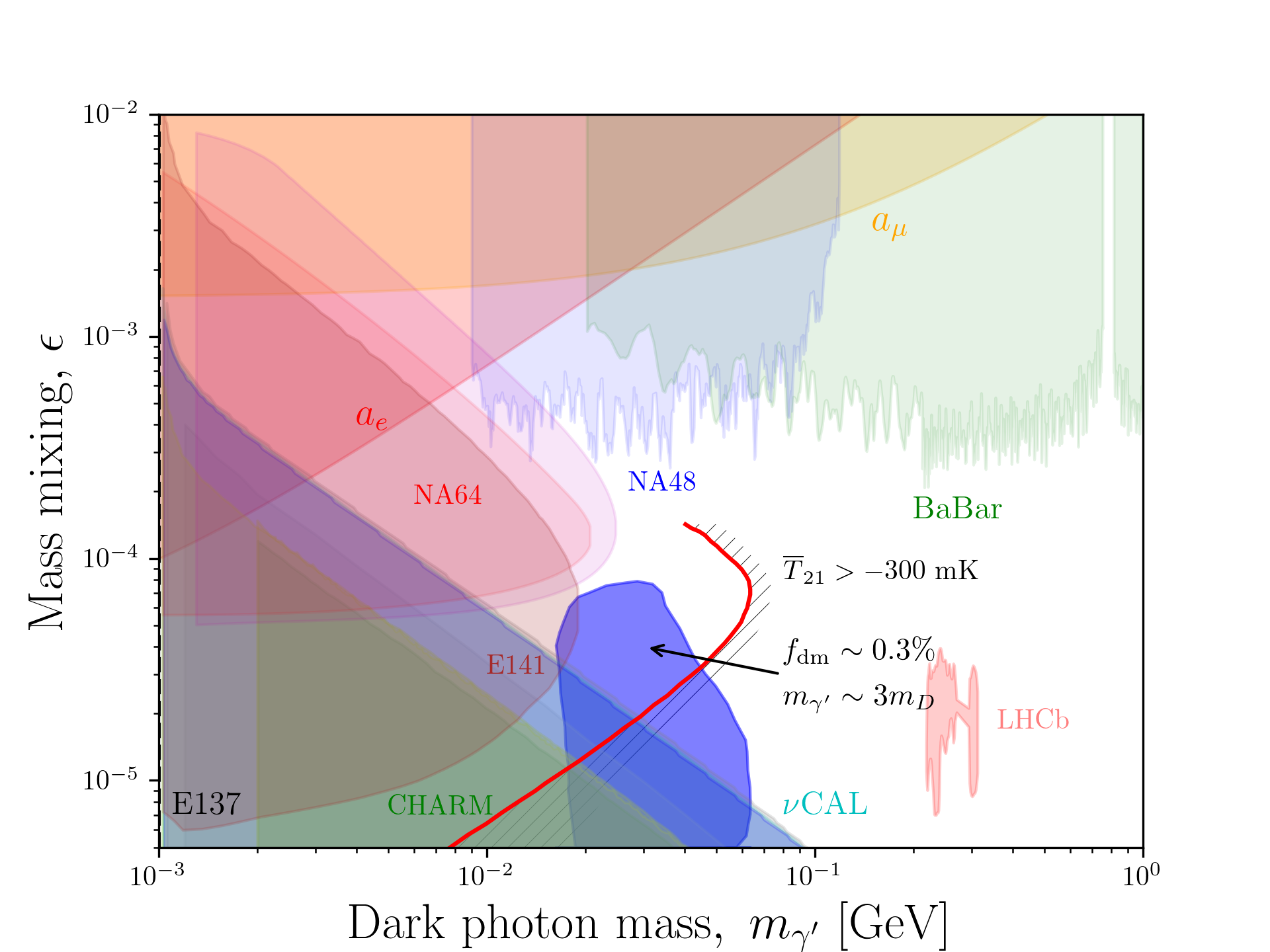}
 \caption{Left panel: Constraints on millicharged dark matter from SLAC~\cite{Prinz:1998ua} (purple region), SN1987A~\cite{Chang:2018rso} (grey region) and CMB limit derived from Planck 2015 data~\cite{Boddy:2018wzy} (pink region). The constraint from EDGES~\cite{Bowman:2018yin} is shown where the excluded red and blue regions correspond to values inconsistent with $\overline{T}_{21}=500^{+200}_{-500}$ mK.
 Also shown are
 contours for fixed $\overline{T}_{21}$ in the allowed region. Right panel: Plot of the mass mixing parameter $\epsilon$ versus the dark photon
 mass where the constraints shown from various experiments are  discussed in the text. The region in blue corresponds to the parameter space giving $f_{\rm dm}\sim 0.3\%$ and is consistent with the allowed region in the $\epsilon_D$-$m_D$ plane
 (shown in the left panel) 
  only within the region not excluded by $\overline{T}_{21}$ bounded by the hatched red curve. Here we set $Q_X=1$ and $g_X=0.2$.  }  
 \label{fig6}
\end{figure}

\section{Conclusion\label{sec:conclu}}

In this work we have presented a particle physics model using the Stueckelberg
mechanism which can generate millicharged dark matter and can provide the
necessary baryon cooling to explain the EDGES data.  The model we consider
is one where the hidden sector possesses a $U(1)$ gauge symmetry and the gauge
field of the hidden sector has a Stueckelberg mass mixing with the
hypercharge gauge field of the standard model. In this case a dark fermion in the hidden sector
develops a millicharge due to the Stueckelberg mass mixing mechanism
 and
 interacts with the baryons via electromagnetic interactions.
We solve a coupled set of equations for the yields of the dark fermion and the
 dark photon, and for the temperatures  of millicharged dark matter and of the baryons as a function
 of redshift.  This allows us to compute the millicharged DM relic density as well as the baryon cooling  as a function of the redshift.  We give a set of benchmarks which
 simultaneously satisfy the required fraction of the millicharged dark matter relic density and of baryon cooling  and thus satisfy the EDGES data and lie in the region of the parameter space
 consistent with all other experimental constraints. \\

\noindent
{\bf Acknowledgments:}  Communications with Ely Kovetz, Hongwan Liu, Julian Mun\~{o}z, and Michael Williams (related to LHCb data on dark
photon limits) 
are acknowledged. 
The research of AA was supported by the BMBF under contract 05H18PMCC1, while the research of PN and ZYW was supported in part by the NSF Grant PHY-1913328.

\section{Appendix \label{sec:append}}
In this appendix we give further details of the analysis presented in the main body of the paper.
Thus in Appendix~\ref{app:A} we briefly describe the underlying canonical normalization of the Lagrangian of Eq.~(\ref{totL}) and 
 display the full neutral current which contains the millicharge coupling. In Appendix~\ref{app:B}, we discuss the cross-sections that enter in the analysis 
 of the relic density of the millicharged dark matter which is constituted of $D$ fermions. The relevant cross sections consist of: (1) $D\bar D \to Z/\gamma/\gamma'\to
 q\bar q$, (2) $D\bar D \to Z/\gamma/\gamma'\to
 \ell\bar \ell$, (3) $D\bar D \to Z/\gamma/\gamma'\to \nu \bar \nu$, (4) $D\bar D \to \gamma' \gamma'$, (5) $q\bar q \to \gamma'$, (6) $\ell \bar \ell \to \gamma'$, 
 (7) $\nu\bar\nu \to \gamma'$, (8) $D\bar D \to \gamma'$, (9) $D\bar D \to \phi \gamma$,  (10) $D\bar D \to \phi \gamma'$, (11) $\gamma' D\to \gamma D$, (12) 
 $\phi D\to \gamma D$ and (13) $D\bar D\to \phi \phi$. In Appendix~\ref{app:C} we discuss the decay of the dark photon which allows the satisfaction of the
 constraint that it decays before BBN.

\appendix

\section{Model details}\label{app:A}

As discussed in section~\ref{sec:model}, in the Standard Model the neutral gauge boson sector comprises of the hypercharge gauge boson
$B^\mu$ and the component $A^\mu_{3}$ of the $SU(2)_L$ gauge field $A^\mu_a$ ($a=$ 1$-$3)
which leads to a $2\times 2$ mass square matrix after spontaneous breaking which contains
one massless mode, the photon, and a massive mode, the $Z$-boson. Inclusion of the Stueckelberg
gauge field $C_\mu$,  enlarges the $2\times 2$ mass square matrix of the
neutral gauge boson sector of the Standard Model to  a $3\times 3$ mass square matrix in the
Stueckelberg extended model. Including the kinetic mixing of the $U(1)_X$ and $U(1)_Y$
and after the transformation to diagonalize the kinetic and mass matrices one has the
 mass eigenstates which are $Z_\mu, A_\mu^{\gamma'}, A_\mu^\gamma$ which are the $Z$-boson, the dark photon
 $\gamma'$ and the photon $\gamma$. 
The couplings of $Z_\mu$ and $A_\mu^{\gamma}$ to the dark fermion were given in Eq.~(\ref{milli-hid1})$-$Eq.~(\ref{milli-hid2})
under the approximation $\epsilon<<1, \delta <<1$. Here we display the full form of the couplings that enter in $\mathcal{L}_{\rm hid} ^{\rm m}$ and 
$\mathcal{L}_{D} ^{\gamma'}$. 
Thus the full forms of the couplings $\epsilon_{Z}^D , \epsilon_{\gamma}^D$ and $g^D_{\gamma'}$ are given by 
\begin{align}
 \epsilon_Z^D &= g_X Q_X (\mathcal{R}_{12}- s_{\delta} \mathcal{R}_{22}),\non 
  \epsilon_{\gamma}^D &= g_X Q_X (\mathcal{R}_{13}- s_{\delta} \mathcal{R}_{23}),\non 
  g_{\gamma'}^D&= g_X Q_X (\mathcal{R}_{11}- s_{\delta} \mathcal{R}_{21}),
 \end{align}
 where $s_{\delta}=\sinh\delta$ and where the matrix $\mathcal{R}$ is given  by Eq.~(23) of~\cite{Feldman:2007wj}. It involves three Euler angles $(\theta, \phi, \psi)$ 
 which diagonalize the Stueckerberg mass  matrix such that $\mathcal{R}^T\mathcal{M}^2 \mathcal {R}= \text{diag}(m^2_{\gamma'},  m_Z^2, 0)$,
 where $\mathcal{M}^2$ is defined by Eq.~(21) of~\cite{Feldman:2007wj}.
The angles $\phi, \theta, \psi$ are defined as
\begin{equation}
 \tan\phi=\bar\epsilon, ~~~ \tan\theta=\frac{g_Y}{g_2}c_{\delta}\cos\phi,
~\tan2\psi=\frac{2\bar\epsilon m^2_Z\sin\theta}{m^2_{\gamma'}-m^2_Z+(m^2_{\gamma'}+m^2_Z-m^2_W)\bar\epsilon^2},
 \label{hid-exact}
\end{equation}
 and $\bar\epsilon=\epsilon c_{\delta}-s_{\delta}$, where $c_{\delta} = \cosh\delta$.
  Under the approximation $\epsilon<<1, \delta <<1$, Eq.~(\ref{hid-exact}) gives the result of Eq.~(\ref{milli-hid1}) and Eq.~(\ref{milli-hid2}). 
  
    As noted earlier, the Standard Model fermions develop millicharge couplings with the dark photon after canonical diagonalization and we get~\cite{Feldman:2007wj}    
 \begin{equation}
\mathcal{L}^{\rm m}_{\rm SM}=\frac{g_2}{2\cos\theta}\bar\psi_f\gamma^{\mu}\Big[(v'_f-\gamma_5 a'_f)A^{\gamma'}_{\mu}\Big]\psi_f,
\label{milli-sm}
\end{equation}
where $f$ runs over all SM fermions and the vector and axial couplings are given by
\begin{equation}
\begin{aligned}
v'_f&=-\cos\psi[(\tan\psi+\bar\epsilon\sin\theta)T_{3f}-2\sin^2\theta(\bar\epsilon\csc\theta+\tan\psi)Q_f],\\
a'_f&=-\cos\psi(\tan\psi+\bar\epsilon\sin\theta)T_{3f},
\end{aligned}
\label{eqn:v-a}
\end{equation}
where  $T_{3f}$ is the third component of isospin and $Q_f$ is the electric charge for the fermion.
 We also record here the couplings of the $Z_\mu$ and $A^\gamma_\mu$ in the canonically diagonalized basis which are given by~\cite{Feldman:2007wj}
\begin{equation}
\mathcal{L}_{\rm SM}=\frac{g_2}{2\cos\theta}\bar\psi_f\gamma^{\mu}\Big[(v_f-\gamma_5 a_f)Z_{\mu}\Big]\psi_f+e\bar\psi_f\gamma^{\mu}Q_f A_{\mu}\psi_f,
\label{SMLag}
\end{equation}
where $f$ runs over all SM fermions and the vector and axial couplings are given by
\begin{equation}
\begin{aligned}
v_f&=\cos\psi[(1-\bar\epsilon\tan\psi\sin\theta)T_{3f}-2\sin^2\theta(1-\bar\epsilon\csc\theta\tan\psi)Q_f],\\
a_f&=\cos\psi(1-\bar\epsilon\tan\psi\sin\theta)T_{3f}. 
\end{aligned}
\label{eqn:v-a}
\end{equation}
In summary, the full set of millicharge couplings arise from Eqs.~(\ref{milli-tot}), (\ref{milli-hid1}),  and~(\ref{milli-sm}). 
Further, the coupling of the dark fermion with the dark photon is normal strength and is given by Eq.~(\ref{D-darkphoton}).

\section{Cross-sections for dark matter relic density calculation}\label{app:B}

We present here the relevant cross-sections needed for the computation of the
millicharged dark matter relic density. The analysis involves all the couplings discussed in section~\ref{sec:model} and in Appendix~\ref{app:A},
specifically Eqs.~(\ref{milli-tot}),  (\ref{milli-hid1}), (\ref{milli-sm}) and~(\ref{D-darkphoton}). 
 These cross sections enter in Eqs.~(\ref{y1})$-$(\ref{y4}).

\begin{enumerate}

\item $D\bar{D}\to Z/\gamma/\gamma' \to q\bar{q}$: \\
The total cross-section for the process $D\bar{D}\to Z/\gamma/\gamma' \to q\bar{q}$ is given by
\begin{align}
\sigma^{D\bar{D}\to q\bar{q}}(s)=&\frac{g_X^2 g_2^2}{8\pi s\cos^2\theta}\sqrt{\frac{s-4m^2_q}{s-4m^2_D}}\Bigg[\frac{(\mathcal{R}_{12}-s_{\delta}\mathcal{R}_{22})^2(\alpha^2\eta_q T^2_{3q}-2\alpha\beta\kappa_q Q_q T_{3q}+2\beta^2 Q_q^2 \kappa_q)}{(s-m_Z^2)^2+m^2_Z\Gamma_Z^2} \nonumber \\
&+\frac{(\mathcal{R}_{11}-s_{\delta}\mathcal{R}_{21})^2(\alpha'^2\eta_q T^2_{3q}-2\alpha'\beta'\kappa_q  Q_q T_{3q}+2\beta'^2 Q_q^2 \kappa_q))}{(s-m^2_{\gamma'})^2+m^2_{\gamma'}\Gamma_{\gamma'}^2} \nonumber \\
&-2(\mathcal{R}_{11}-s_{\delta}\mathcal{R}_{21})(\mathcal{R}_{12}-s_{\delta}\mathcal{R}_{22})\Big\{Q_q\beta (2\beta' Q_q-\alpha' T_{3q})\kappa_q \nonumber \\
&+\alpha T_{3q}(\alpha' T_{3q}\eta_q-\beta' Q_q \kappa_q)\Big\}
\times\frac{(s-m^2_Z)(s-m^2_{\gamma'})+\Gamma_Z\Gamma_{\gamma'}m_Z m_{\gamma'}}{[(s-m_Z^2)^2+m^2_Z\Gamma_Z^2][(s-m^2_{\gamma'})^2+m^2_{\gamma'}\Gamma_{\gamma'}^2]}\Bigg] \nonumber \\
&+\frac{g_X^2}{4\pi s}\sqrt{\frac{s-4m_q^2}{s-4m_D^2}}\Bigg[\frac{e^2 Q_q^2\kappa_q}{s^2}(\mathcal{R}_{13}-s_{\delta}\mathcal{R}_{23})^2-\frac{e g_2 Q_q\kappa_q}{\cos\theta}\times \nonumber \\
&\Bigg\{\frac{s-m^2_Z}{s[(s-m_Z^2)^2+m^2_Z\Gamma^2_Z]}(2\beta Q_q-\alpha T_{3q})(\mathcal{R}_{12}-s_{\delta}\mathcal{R}_{22})(\mathcal{R}_{13}-s_{\delta}\mathcal{R}_{23}) \nonumber \\
&-\frac{s-m^2_{\gamma'}}{s[(s-m_{\gamma'}^2)^2+m^2_{\gamma'}\Gamma^2_{\gamma'}]}(2\beta' Q_q-\alpha' T_{3q})(\mathcal{R}_{11}-s_{\delta}\mathcal{R}_{21})(\mathcal{R}_{13}-s_{\delta}\mathcal{R}_{23})\Bigg\}\Bigg],
\end{align}
where $e=\gamma\cos\theta$, $m_q$, $m_Z$ and $m_{\gamma'}$ are the quark, $Z$ and $\gamma'$ masses, respectively, and $T_{3q}=1/2 (-1/2)$ and $Q_q=2/3 (-1/3)$ for up-(down)-type quarks, and with
\begin{equation}
\begin{aligned}
\kappa_q&=(s+2m^2_D)(s+2m^2_q),~~~\eta_q=(s+2m^2_D)(s-m^2_q), \\
\alpha&=\cos\psi-\bar{\epsilon}\sin\theta\sin\psi,~~~\beta=\sin^2\theta\cos\psi-\bar{\epsilon}\sin\theta\sin\psi, \\
\alpha'&=\sin\psi+\bar{\epsilon}\sin\theta\cos\psi,~~~\beta'=\sin^2\theta\sin\psi+\bar{\epsilon}\sin\theta\cos\psi.
\end{aligned}
\end{equation}

\item $D\bar{D}\to Z/\gamma/\gamma'  \to\ell\bar{\ell}$: \\
The total cross-section for the process $D\bar{D}\to Z/\gamma/\gamma' \to\ell\bar{\ell}$ is given by
\begin{align}
\sigma^{D\bar{D}\to\ell\bar{\ell}}(s)=&\frac{g_X^2 g_2^2}{96\pi s\cos^2\theta}\sqrt{\frac{s-4m^2_{\ell}}{s-4m^2_D}}\Bigg[\frac{(\mathcal{R}_{12}-s_{\delta}\mathcal{R}_{22})^2(\alpha^2\eta_{\ell}-4\alpha\beta\kappa_{\ell} +8\beta^2 \kappa_{\ell})}{(s-m_Z^2)^2+m^2_Z\Gamma_Z^2} \nonumber \\
&+\frac{(\mathcal{R}_{11}-s_{\delta}\mathcal{R}_{21})^2(\alpha'^2\eta_{\ell}-4\alpha'\beta'\kappa_{\ell} +8\beta'^2 \kappa_{\ell})}{(s-m_{\gamma'}^2)^2+m^2_{\gamma'}\Gamma_{\gamma'}^2} \nonumber \\
&+2(\mathcal{R}_{11}-s_{\delta}\mathcal{R}_{21})(\mathcal{R}_{12}-s_{\delta}\mathcal{R}_{22})\Big\{2\beta(\alpha'-4\beta')\kappa_{\ell} \nonumber \\
&-\alpha (\alpha'\eta_{\ell}-2\beta' \kappa_{\ell})\Big\}\times      \frac{(s-m^2_Z)(s-m^2_{Z'})+\Gamma_Z\Gamma_{\gamma'}m_Z m_{\gamma'}}{[(s-m_Z^2)^2+m^2_Z\Gamma_Z^2][(s-m^2_{\gamma'})^2+m^2_{\gamma'}\Gamma_{\gamma'}^2]} \Bigg] \nonumber \\
&+\frac{g_X^2}{24\pi s}\sqrt{\frac{s-4m_{\ell}^2}{s-4m_D^2}}\Bigg[\frac{2e^2\kappa_{\ell}}{s^2}(\mathcal{R}_{13}-s_{\delta}\mathcal{R}_{23})^2+\frac{e g_2\kappa_{\ell}}{\cos\theta}\times \nonumber \\
&\Bigg\{\frac{s-m^2_Z}{s[(s-m_Z^2)^2+m^2_Z\Gamma^2_Z]}(\alpha-4\beta)(\mathcal{R}_{12}-s_{\delta}\mathcal{R}_{22})(\mathcal{R}_{13}-s_{\delta}\mathcal{R}_{23}) \nonumber \\
&-\frac{s-m^2_{\gamma'}}{s[(s-m_{\gamma'}^2)^2+m^2_{\gamma'}\Gamma^2_{\gamma'}]}(\alpha'-4\beta')(\mathcal{R}_{11}-s_{\delta}\mathcal{R}_{21})(\mathcal{R}_{13}-s_{\delta}\mathcal{R}_{23})\Bigg\}\Bigg],
\end{align}
where
\begin{align}
\kappa_{\ell}&=(s+2m^2_D)(s+2m^2_{\ell}),~~~\eta_{\ell}=(s+2m^2_D)(s-m^2_{\ell}).
\end{align}

\item $D\bar{D}\to Z/\gamma' \to\nu\bar{\nu}$: \\
The total cross-section for the process $D\bar{D}\to Z/\gamma' \to\nu\bar{\nu}$ is given by
\begin{align}
\sigma^{D\bar{D}\to\nu\bar{\nu}}(s)&=\frac{g_X^2 g^2_2 }{32\pi\cos^2\theta}\frac{(s+2m^2_D)s^{1/2}}{\sqrt{s-4m^2_D}}\Bigg\{\frac{\alpha'^2(\mathcal{R}_{11}-s_{\delta}\mathcal{R}_{21})^2}{(s-m_{\gamma'}^2)^2+m^2_{\gamma'}\Gamma_{\gamma'}^2}+\frac{\alpha^2(\mathcal{R}_{12}-s_{\delta}\mathcal{R}_{22})^2}{(s-m_{Z}^2)^2+m^2_{Z}\Gamma_{Z}^2} \nonumber \\
&-2\alpha\alpha'(\mathcal{R}_{11}-s_{\delta}\mathcal{R}_{21})(\mathcal{R}_{12}-s_{\delta}\mathcal{R}_{22}) \nonumber \\
&\times\frac{(s-m^2_Z)(s-m^2_{\gamma'})+\Gamma_Z\Gamma_{\gamma'}m_Z m_{\gamma'}}{[(s-m_Z^2)^2+m^2_Z\Gamma_Z^2][(s-m^2_{\gamma'})^2+m^2_{\gamma'}\Gamma_{\gamma'}^2]} \Bigg\}.
\end{align}

\item $D\bar{D}\longleftrightarrow \gamma' \gamma'$: \\
The total cross-section for the process $D\bar{D}\to \gamma' \gamma'$ is
\begin{align}
\sigma^{D\bar{D}\to \gamma' \gamma'}(s)=\frac{g_{X}^4(\mathcal{R}_{11}-s_{\delta}\mathcal{R}_{21})^4}{8\pi s(s-4m^2_D)}&\Bigg\{-\frac{\sqrt{(s-4m^2_{\gamma'})(s-4m^2_{D})}}{m^4_{\gamma'}+m^2_D(s-4m^2_{\gamma'})}[2m^4_{\gamma'}+m^2_D(s+4m^2_D)] \nonumber \\
&+\frac{\log A}{s-2m^2_{\gamma'}}(s^2+4m^2_D s+4m^4_{\gamma'}-8m^4_D-8m_D^2 m^2_{\gamma'})\Bigg\},
\end{align}
where
\begin{equation}
A=\frac{s-2m^2_{\gamma'}+\sqrt{(s-4m^2_{\gamma'})(s-4m^2_{D})}}{s-2m^2_{\gamma'}-\sqrt{(s-4m^2_{\gamma'})(s-4m^2_{D})}},
\end{equation}

The reverse processes are given by
\begin{equation}
9(s-4m^2_{\gamma'})\sigma^{\gamma' \gamma'\to D\bar{D}}(s)=8(s-4m^2_{D})\sigma^{D\bar{D}\to \gamma'\gamma'}(s).
\end{equation}

\item $q\bar{q}\to \gamma'$: \\

The cross-section for the process $q\bar{q}\to \gamma'$ is
\begin{align}
\sigma^{q\bar{q}\to \gamma'}(s)&=\frac{\pi(g_2^2+\gamma^2)}{6s(s-4m_q^2)^{1/2}}\Bigg[2Q_q^2\beta'^2(m_{\gamma'}^2+2m^2_q)-2Q_qT_{3q}\alpha'\beta'(m_{\gamma'}^2+2m^2_q) \nonumber \\
&\hspace{3.5cm}+\alpha'^2 T_{3q}^2(m_{\gamma'}^2-m^2_q)\Bigg]\delta(\sqrt{s}-m_{\gamma'}).
\end{align}

\item $\ell\bar{\ell}\to \gamma'$: \\

The cross-section for the process $\ell\bar{\ell}\to \gamma'$ is
\begin{align}
\sigma^{\ell\bar{\ell}\to \gamma'}(s)=\frac{\pi(g_2^2+\gamma^2)}{2s(s-4m_{\ell}^2)^{1/2}}&[2\beta'^2(m_{\gamma'}^2+2m^2_{\ell})-\alpha'\beta'(m_{\gamma'}^2+2m^2_{\ell}) \nonumber \\
&+\frac{1}{4}\alpha'^2(m_{\gamma'}^2-m^2_{\ell})]\delta(\sqrt{s}-m_{\gamma'}).
\end{align}

\item $\nu\bar{\nu}\to \gamma'$: \\

The cross-section for the process $\nu\bar{\nu}\to \gamma'$ is
\begin{equation}
\sigma^{\nu\bar{\nu}\to \gamma'}(s)=\frac{3\pi(g_2^2+\gamma^2)\alpha'^2 m^2_{\gamma'}}{4s^{3/2}}\delta(\sqrt{s}-m_{\gamma'}).
\end{equation}

\item $D\bar{D}\to\gamma'$: \\

The cross-section for the process $D\bar{D}\to \gamma'$ is
\begin{equation}
\sigma^{D\bar{D}\to \gamma'}(s)=\frac{\pi g_X^2}{2s^{3/2}}(\mathcal{R}_{11}-s_{\delta}\mathcal{R}_{21})^2(2m^2_D+m_{\gamma'}^2)\delta(\sqrt{s}-m_{\gamma'}).
\end{equation}

\item $D\bar{D}\to \phi\gamma$: \\

The cross-section for the process $D\bar{D}\to \phi\gamma$ is 
\begin{align}
\sigma^{D\bar{D}\to \phi\gamma}(s)&=\frac{g_X^2\lambda^2(\mathcal{R}_{13}-s_{\delta}\mathcal{R}_{23})^2}{8\pi s^2(s-4m_D^2)}\left[8m_D^2 s\sqrt{1-\frac{4m_D^2}{s}}+(s-4m_D^2)^2\log B\right],
\end{align}
where 
\begin{equation}
B=\frac{s-2m_D^2+\sqrt{s(s-4m_D^2)}}{2m_D^2}.
\end{equation}

\item $D\bar{D}\to \phi\gamma^\prime$: \\

The cross-section for the process $D\bar{D}\to \phi\gamma^\prime$ is
\begin{align}
\sigma^{D\bar{D}\to \phi\gamma^\prime}(s)&=\frac{g_X^2\lambda^2(\mathcal{R}_{11}-s_{\delta}\mathcal{R}_{21})^2}{8\pi s(s-4m_D^2)(s-m_{\gamma'}^2)}\Bigg\{4(2m_D^2+m_{\gamma'}^2)\sqrt{s(s-4m_D^2)} \nonumber \\
&\hspace{2cm}+[16m_D^4+(s-m_{\gamma'}^2)^2-8m_D^2(s-2m_{\gamma}^2)]\log B\Bigg\}.
\end{align}

\item $\gamma' D\to \gamma D$: \\

The cross-section for the process $\gamma' D\to \gamma D$ is
\begin{align}
\sigma^{\gamma' D\to\gamma D}(s)&=\frac{g_X^4}{24\pi s^2(s-m_D^2)}\frac{(\mathcal{R}_{11}-s_{\delta}\mathcal{R}_{21})^2(\mathcal{R}_{13}-s_{\delta}\mathcal{R}_{23})^2}{m_D^4+(s-m_{\gamma'}^2)^2-2m_D^2(s+m_{\gamma}^2)} \nonumber \\
&\times\Bigg\{\sqrt{m_D^4+(s-m_{\gamma'}^2)^2-2m_D^2(s+m_{\gamma}^2)}\,[m_D^6-m_D^4(s+m_{\gamma'}^2) \nonumber \\
&~~~+s^2(s+7m_{\gamma'}^2)+m_D^2 s(15s+2m_{\gamma'}^2)]+2[3m_D^4-2m_{\gamma'}^4 \nonumber \\
&~~~+2m_D^2(3s-m_{\gamma'}^2)+2s(m_{\gamma'}^2-s)]\log C\Bigg\},
\end{align} 
where
\begin{equation}
C=\frac{m_D^2(m_D^2-m_{\gamma'}^2)+s(m_{\gamma'}^2-s)+(s-m_D^2)\sqrt{m_D^4+(s-m_{\gamma'}^2)^2-2m_D^2(s+m_{\gamma}^2)}}{m_D^2(m_D^2-m_{\gamma'}^2)+s(m_{\gamma'}^2-s)-(s-m_D^2)\sqrt{m_D^4+(s-m_{\gamma'}^2)^2-2m_D^2(s+m_{\gamma}^2)}}.
\end{equation}

\item $\phi D\to \gamma D$: \\

The cross-section for the process $\phi D\to \gamma D$ is
\begin{align}
\sigma^{\phi D\to \gamma D}(s)&=\frac{g_X^2\lambda^2(\mathcal{R}_{13}-s_{\delta}\mathcal{R}_{23})^2}{16\pi s^2(s-m_D^2)^2}\Bigg[(3s+m_D^2)(m_D^4-8s m_D^2-s^2) \nonumber \\
&\hspace{4cm}-2s^2\frac{(s+3m_D^2)^2}{s-m_D^2}\log\left(\frac{m_D^2}{s}\right)\Bigg].
\end{align}

\item $D\bar{D}\to \phi\phi$: \\

The cross-section for the process $D\bar{D}\to \phi\phi$ is
\begin{align}
\sigma^{D\bar{D}\to \phi\phi}(s)&=\frac{\lambda^4}{32\pi s^2(s-4m_D^2)}\Bigg[\frac{2(s-m_D^2)^2(m_D^6-4m_D^4 s-26 m_D^2 s^2-4s^3)}{(2s-m_D^2)(m_D^4-2m_D^2 s+2s^2)} \nonumber \\
&+(32m_D^4-16m_D^2 s-s^2)\log\left(\frac{2m_D^2 s-m_D^4}{2s^2-2m_D^2 s+m_D^4}\right)\Bigg].
\end{align}

\end{enumerate}

\section{Dark photon decay}\label{app:C}

\begin{enumerate}

\item The decay width of $\gamma'$ to leptons is given by
\begin{align}
\Gamma_{\gamma'\to\ell\bar\ell}=\frac{g_2^2}{24\pi m_{\gamma'}\cos^2\theta}\sqrt{1-\left(\frac{2m_{\ell}}{m_{\gamma'}}\right)^2}&\Bigg[\frac{1}{4}\alpha'^2(m^2_{\gamma'}-m^2_{\ell})-\alpha'\beta'(m^2_{\gamma'}+2m^2_{\ell})\nonumber \\
&+2\beta'^2(m^2_{\gamma'}+2m^2_{\ell})\Bigg].
\end{align}

\item The decay width of $\gamma'$ to quarks is given by
\begin{align}
\Gamma_{\gamma'\to q\bar q}=\frac{g_2^2}{8\pi m_{\gamma'}\cos^2\theta}\sqrt{1-\left(\frac{2m_{q}}{m_{\gamma'}}\right)^2}&\Bigg[\alpha'^2(m^2_{\gamma'}-m^2_{q})T_{3q}^2-2\alpha'\beta'(m^2_{\gamma'}+2m^2_{q})Q_q T_{3q}\nonumber \\
&+2\beta'^2(m^2_{\gamma'}+2m^2_{q})Q_q^2\Bigg].
\end{align}

\item The decay width of $\gamma'$ to neutrinos is given by
\begin{equation}
\Gamma_{\gamma'\to\nu\bar\nu}=\frac{g_2^2}{32\pi\cos^2\theta}m_{\gamma'}\alpha'^2.
\end{equation}

\item The decay width of $\gamma'$ to dark fermions is given by
\begin{equation}
\Gamma_{\gamma'\to D\bar{D}}=\frac{g_X^2}{12\pi}(\mathcal{R}_{11}-s_{\delta}\mathcal{R}_{21})^2 m_{\gamma'}\left(1+\frac{2m^2_D}{m^2_{\gamma'}}\right).
\end{equation} 

\end{enumerate}

\end{document}